\documentclass[final,3p,times,twocolumn]{elsarticle}

\usepackage{amssymb}
\usepackage{xcolor}
\usepackage{amsmath}
\usepackage{mhchem}
\usepackage{booktabs}
\usepackage{multirow}
\usepackage{ulem}
\usepackage{xcolor}    
\usepackage{xspace}    
\usepackage{graphicx}
\usepackage{xcolor}
\definecolor{omittext}{RGB}{0,0,0}  

\newcommand{\strikeout}[1]{\unskip}  

\journal{Solar Energy 296, 113535 (2025)}
\usepackage{hyperref}
\usepackage{etoolbox}
\makeatletter
\patchcmd{\ps@pprintTitle}
  {Preprint submitted to}
  {DOI: \textbf{\href{https://doi.org/10.1016/j.solener.2025.113535}{10.1016/j.solener.2025.113535} }}
  {}{}
\makeatother

\begin{document}

\begin{frontmatter}




\title{Efficiency Enhancement of c-Si/\ce{TiO2} Heterojunction Thin Film Solar Cell Using Hybrid Metal-Dielectric Nanostructures}


\author[lev1]{Soikot Sarkar} 
\author[lev1]{Sajid Muhaimin Choudhury\corref{cor1}}

\ead{sajid@eee.buet.ac.bd}
\cortext[cor1]{Corresponding Author.}
\affiliation[lev1]{organization={Department of Electrical and Electronic Engineering, Bangladesh University of Engineering and Technology},
            city={Dhaka},
            postcode={1205},
            country={Bangladesh}}

\begin{abstract}
The hybrid metal-dielectric nanostructures (HMDN) are promising candidates to address the ohmic loss by conventional nanostructures in photovoltaic applications by strong confinement and high scattering directivity. In this study, we present a c-\ce{Si}/\ce{TiO2} heterojunction thin film solar cell (TFSC) where a pair of triangular HMDN comprised of \ce{Ag} and \ce{AZO} was utilized to enhance the longer wavelength light absorption. The presence of the \ce{TiO2} inverted pyramid layer, in combination with the \ce{ITO} and \ce{SiO2}-based pyramid layers at the front, enhanced the shorter wavelength light absorption by increasing the optical path and facilitating the coupling of incoming light in photonic mode. Consequently, the average absorption by 1000 nm thick photoactive layer reached 83.32 \% for AM 1.5G within the wavelength range of 300 – 1100 nm which was investigated by employing the finite-difference time-domain (FDTD) method.  The electric field profile and absorbed power density profile demonstrated the respective contributions of each layer in the absorption of light at shorter and longer wavelengths. The structure exhibited a short circuit current density ($J_{sc}$) of 37.96 mA/cm$^2$ and a power conversion efficiency ($PCE$) of 17.42 \%. The efficiency of our proposed structure experienced a maximum relative change of 0.34 \% when a polarized light was exposed with an angle of 0$^\circ$ to 90$^\circ$. The incorporation of self-heating in non-isothermal conditions reduced $PCE$ by 13.77 \%.  In addition, the comparative analysis to assess the impact of HMDN on our structure revealed a 4.54 \% increase in $PCE$ of the structure with metallic nanostructures, paving the way for the utilization of HMDN to enhance the performance of TFSC.
\end{abstract}

\begin{keyword}
c-\ce{Si}/\ce{TiO2} heterojunction \sep thin film solar cell \sep Hybrid metal-dielectric nanostructure \sep Polarization angle tolerant \sep FDTD \sep Surface plasmon resonance 

\end{keyword}

\end{frontmatter}



\section{Introduction}\label{sec1}
Silicon solar cells, along with perovskite, \ce{GaAs}, CIGS, \ce{CdTe} based solar cells, have been the promising technology in the photovoltaic industry for many years, propelling the world towards renewable energy~\cite{Qin2023, He2024, Patel2023, Zheng2023, Pokhrel2022}. However, crystalline Si (c-Si)-based solar cells prevail in the photovoltaic industry due to the abundance of \ce{Si} material availability and well-established fabrication techniques. \ce{Si} exhibited lower photo-conversion efficiency due to its indirect bandgap. Therefore, a thick layer of Si photoactive layer (around $100\mu m$) is required to capture sufficient light for improved power conversion efficiency (PCE)~\cite{Zhao2021, Sun2022, Ballif2022}. Although the presence of the thick Si photoactive layer exhibits improved optical performance, it poses high material consumption. Moreover, the thick photoactive layer promotes significant recombination of photo-generated charge carriers that can not contribute to the photocurrent. In this regard, TFSC has garnered the attention of researchers due to its low material consumption as well as low bulk recombination properties. This low bulk recombination feature allows TFSC to be compatible with high carrier collecting capabilities as well as produce high open circuit voltage. However, the reduced photoactive area of TFSC allows for a lower absorption of incoming photons, resulting in a smaller current density. To enhance the current density, various light-trapping mechanisms have been developed in recent years, including nanoholes~\cite{Moeini2023}, nanowires~\cite{El-Bashar2022, Khaled2020}, nanocones~\cite{Wang2022, Prashant2022}, trapezoid pyramids~\cite{Zhao2022, Oliveto2021}, and plasmonic nanostructures (NS)~\cite{Ibrahim2021, Jamil2023, Rassekh2021}. Moreover, the presence of a textured structure at the front significantly reduces the reflection of incident light, hence enhancing the absorption of light by the photoactive layer~\cite{Lozac'h2020, Meyer2021}. Currently, researchers are focusing on plasmonic NS because they have the improved ability to enhance light absorption through surface plasmon resonance (SPR). The plasmonic structure enhances the optical absorption by scattering and coupling the incoming light in SPR mode~\cite{Pritom2023}. The scattering effect improves the optical path length within the photoactive layer, hence enhancing light absorption. By adjusting the structural shape, material types, and surrounding medium of the NS, it is conceivable to modify both the resonance wavelength and the intensity of the resonance~\cite{Cui2020, Ibrahim2021, Alkhalayfeh2021}. The conventional method for constructing the plasmonic NS involves adorning the metal nanoparticles. However, due to the significant ohmic losses demonstrated by the NS comprised of metal, a portion of the incoming light is transformed into thermal energy. The incoming photons are directly absorbed within the metal rather than the semiconductor. As a result, these absorbed photons do not contribute to generating electron-hole pairs as well as photocurrent~\cite{Barreda2022}. This limitation reduces the effectiveness of incorporating metal-based NS to enhance light absorption. To address this challenge, HMDN has garnered the attention of researchers due to the minimized parasitic absorption. In HMDN, the metallic part demonstrates strong confinement of electromagnetic waves around it while the dielectric part exhibits high scattering directivity. We can incorporate the NS both at the front and bottom surface of the photoactive layer. Since the longer wavelength photons have the ability to penetrate through the photoactive layer, the NS-based metallic back reflector can be utilized to trap these penetrated photons~\cite{Ren2018}. Moreover, this back reflector serves as a metal contact of the cell. Consequently, a metallic back reflector composed of NS is prominent for TFSC to enhance the photovoltaic performance. In addition, to improve the efficiency of photovoltaic systems, it is crucial to select contact materials that facilitate the separation of electron-hole pairs and minimize recombination. Titanium dioxide (\ce{TiO2}) is frequently used as an effective electron transport layer in a range of solar cell technologies, such as perovskite solar cells, dye-sensitized solar cells, and crystalline silicon (c-\ce{Si}) solar cells~\cite{Mohammed2022, Wang2020, Elsaeedy2021}.\\

In this study, we present a polarization tolerant c-\ce{Si}/\ce{TiO2} heterostructure TFSC where a pair of triangular HMDN comprised of \ce{Ag} and Aluminum doped zinc oxide (\ce{AZO}) were periodically arranged on the Ag back reflector. We investigated the optical performance of different structures under the illumination of incident light (AM 1.5G) within the wavelength range of 300 – 1100 nm where our proposed structure exhibited an average absorption of 83.32 \% under unpolarized light. We also investigated the electrical performance of our proposed structure by utilizing the generation data of the \ce{Si} photoactive region from the optical simulation. The short circuit current density, $J_{sc}$, and open circuit voltage, $V_{oc}$ of our proposed structure were determined to be 37.96 mA/cm$^2$ and 0.56 V. The structure exhibited a power conversion efficiency, $PCE$ of 17.42 \% with a fill factor, $FF$ of 0.82. The contributions of the NS and top layers to enhance the absorption are discussed through the electric field profile and absorbed power density profile. Since polarization-independent capability is required to achieve optimal photovoltaic efficiency, we investigated the optical and electrical performance under polarized light with angles ranging from 0$^\circ$ to 90$^\circ$. The structure exhibited a maximum of 0.34 \% relative change of PCE. Moreover, we investigated the impact of the incident angle ranging from 0$^\circ$ to 60$^\circ$ both on optical and electrical performance. We also performed an investigation on photovoltaic performance by varying the structural parameters. Furthermore, we investigated the electrical performance of the structures under non-isothermal conditions by incorporating self-heating. In addition, we analyzed the electrical performance of the structure by integrating the pair of dielectric NS, metallic NS, and HMDN separately. An extensive comparative analysis of electrical performance was conducted among our proposed structure and previously reported structures.

\section{Structure Design and Methodology}\label{sec2}
\begin{figure*}[h!]
    \centering
    \includegraphics[width = 1\linewidth]{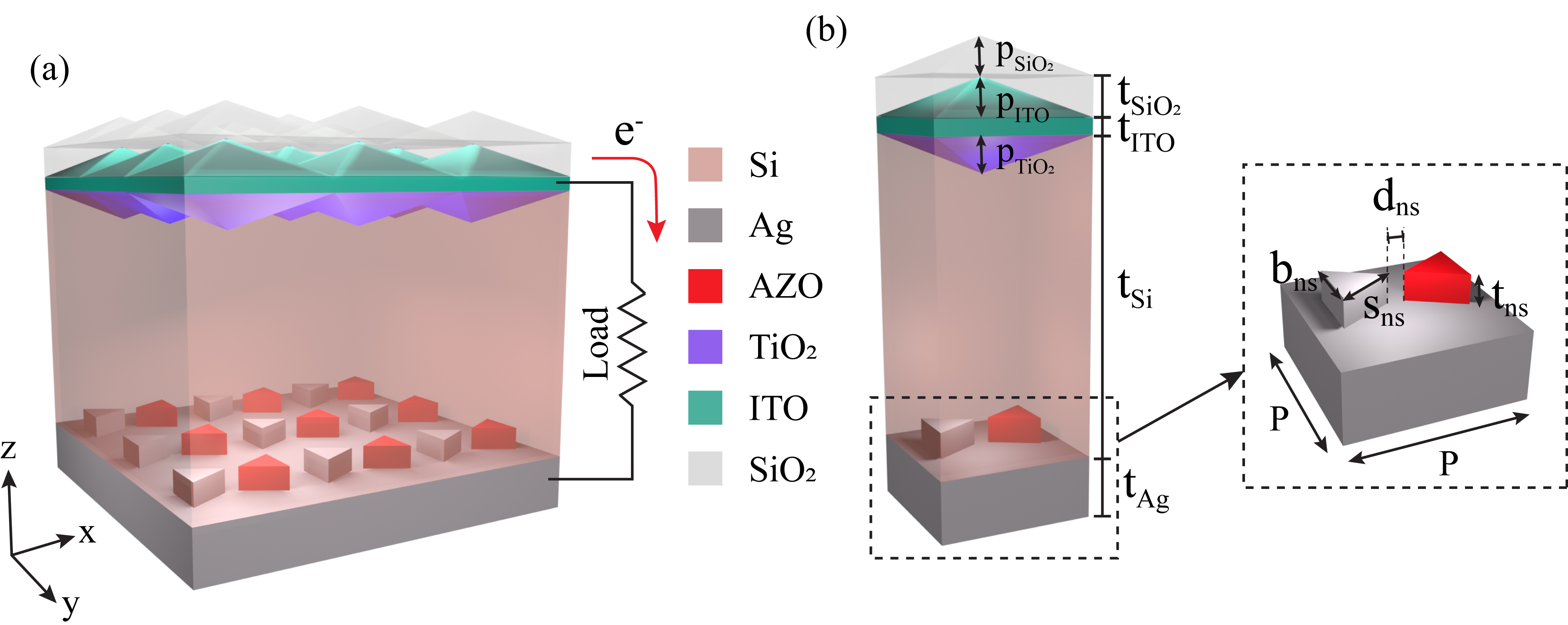}
    \caption{Schematic illustration of our proposed heterojunction TFSC (a) 3D - view (b) 3D - view of a unit cell (Inset shows the zoom view of bottom layer). The structural parameters are considered as, t$_{\ce{Ag}}$ = 100 nm, t$_{\ce{Si}}$ = 1000 nm, t$_{\ce{ITO}}$ = 20 nm, t$_{\ce{SiO2}}$ = 140 nm, p$_{\ce{TiO2}}$ = 80 nm, p$_{\ce{ITO}}$ = 100 nm, p$_{\ce{SiO2}}$ = 100 nm, P = 150 nm, b$_{\ce{ns}}$ = 50 nm, s$_{\ce{ns}}$ = 55 nm, t$_{\ce{ns}}$ = 160 nm, d$_{\ce{ns}}$ = 10 nm.}
    \label{Structure}
\end{figure*}
Figure~\ref{Structure} (a) shows the schematic illustration of our proposed c-\ce{Si}/\ce{TiO2} heterojunction TFSC. Moreover, Fig.~\ref{Structure} (b) depicts a unit cell of our proposed structure with different design parameters where the inset shows a zoomed view of the bottom layer. The designed structure has periodicity both in the x and y directions with the period, P of 150 nm. To achieve an optimum absorption and a negligible transmission of incident light through the photoactive layer, a \ce{Ag} based BR was employed to reflect the penetrated incident light back into the photoactive layer. A negligible amount of transmission through the structure was observed while the thickness of \ce{Ag} slab, t$_{\ce{Ag}}$ of 100 nm was utilized as BR. Moreover, this \ce{Ag} based BR serves the role of the base of our proposed structure. A pair of isosceles triangular HMDN comprised of \ce{Ag} and Aluminium doped Zinc Oxide (\ce{AZO}) were employed on the BR to enhance the absorption of longer wavelengths of light. This pair of HMDN were separated with a distance, d$_{\ce{ns}}$ of 10 nm. The thickness, base, and sides of each triangular NS were considered as t$_{\ce{ns}}$ = 160 nm, b$_{\ce{ns}}$ = 50 nm and s$_{\ce{ns}}$ = 55 nm. 
A c-Si photoactive layer was placed at the upper periphery of the BR which made the position of the NS pair at the interface of the c-Si layer and BR. To minimize material usage and achieve a thin structure, we opted for a c-Si layer thickness, t$_{\ce{Si}}$ of 1000 nm.  We employed an inverted pyramid-shaped structure comprised of \ce{TiO2} with a thickness, p$_{\ce{TiO2}}$ of 80 nm. This inverted pyramid layer played the role of the electron transport layer of our proposed structure. A combination of a planar layer and a pyramid structure comprised of Indium Tin Oxide (\ce{ITO}) was introduced at the surface of the \ce{TiO2} layer. The thickness of the planar layer and the pyramid were considered as t$_{\ce{ITO}}$ = 20 nm, and p$_{\ce{ITO}}$ = 100 nm respectively.  This composed structure served the role of the emitter as well as enhanced the light-capturing capabilities of our proposed TFSC. Here, we used both \ce{TiO2} and \ce{ITO} for being transparent conducting oxide (TCO) material. In addition, a \ce{SiO2} layer with a thickness, t$_{\ce{SiO2}}$ of 140 nm was employed on the upper surface of the \ce{ITO}-based pyramid to enhance the absorption of light and reduce the reflection of incidence light. Consequently, a pyramid of \ce{SiO2} with a thickness, p$_{\ce{SiO2}}$ of 100 nm was created. The values of all structural parameters were optimized by numerical approaches (see \textcolor{blue}{Section S1} of \textcolor{blue}{ESI} for details). The complex refractive indices data of \ce{Si}, \ce{Ag}, and \ce{SiO2} were obtained from Palik \textit{et al.}~\cite{Palik2012}. Moreover, we adopted the data of refractive index and extinction coefficient of \ce{TiO2}, \ce{ITO}, and \ce{AZO} from DeVore \textit{et al.}~\cite{DeVore1951}, K{\"o}nig \textit{et al.}~\cite{König2014}, and Treharne \textit{et al.}~\cite{Treharne2011} respectively (see \textcolor{blue}{Section S2} of \textcolor{blue}{ESI} for details).

In this study, we employed the 3D finite-difference time-domain (FDTD) (Ansys Lumerical) to conduct the optical simulation of our proposed structure. Considering the periodicity observed in the structure in both the x and y directions, we implemented periodic boundary conditions in both of these directions. Additionally, we employed a total of 12 perfectly matched steep-angle layers in the positive and negative z-direction to prevent the phenomenon of parasitic reflection originating from the structure. To achieve an optimal balance between accuracy, memory usage, and simulation time, we employed a non-uniform mesh with a minimum mesh step of 0.25 nm and a mesh accuracy of 3. For mesh refinement, we used conformal variant 0, wherein Conformal Mesh Technology (CMT) was applied to all materials except metals. We utilized CW-normalized transverse magnetic (TM), transverse electric (TE), and unpolarized plane waves with wavelengths ranging from 300 to 1100 nm to be incident from the upper surface of the proposed TFSC. Frequency domain field and power monitors were employed to quantify the amount of light passed through the photoactive layer and the amount of light reflected from the photoactive layer. The monitors captured the characteristics of transmittance, $T(\lambda)$, and reflectance, $R(\lambda)$ as a function of wavelength which are as follows,
\begin{equation}
    T(\lambda) = \frac{P_T(\lambda)}{P_I(\lambda)} , \ce{and}
\end{equation}
\begin{equation}
    R(\lambda) = \frac{P_R(\lambda)}{P_I(\lambda)}
\end{equation}
The absorptance of the photoactive layer was calculated by, 
\begin{equation}
    A(\lambda) = 1 - T(\lambda) - R(\lambda)
\end{equation}
We utilized the unpolarized plane wave to evaluate the generation rate, as our proposed structure exhibits asymmetrical characteristics. However, the polarized plane wave was employed to investigate the impact of polarization angle on electrical performance. After post-processing the generation rate data, we utilized this data as input for the Charge solver to evaluate the electrical performance parameters. Some mathematical approaches were utilized to compute the generation rate (see \textcolor{blue}{Section S3} of \textcolor{blue}{ESI} for details). All the simulations stated above were carried out at a temperature of 300 K.

\begin{figure}
    \centering
    \includegraphics[width = 1\linewidth]{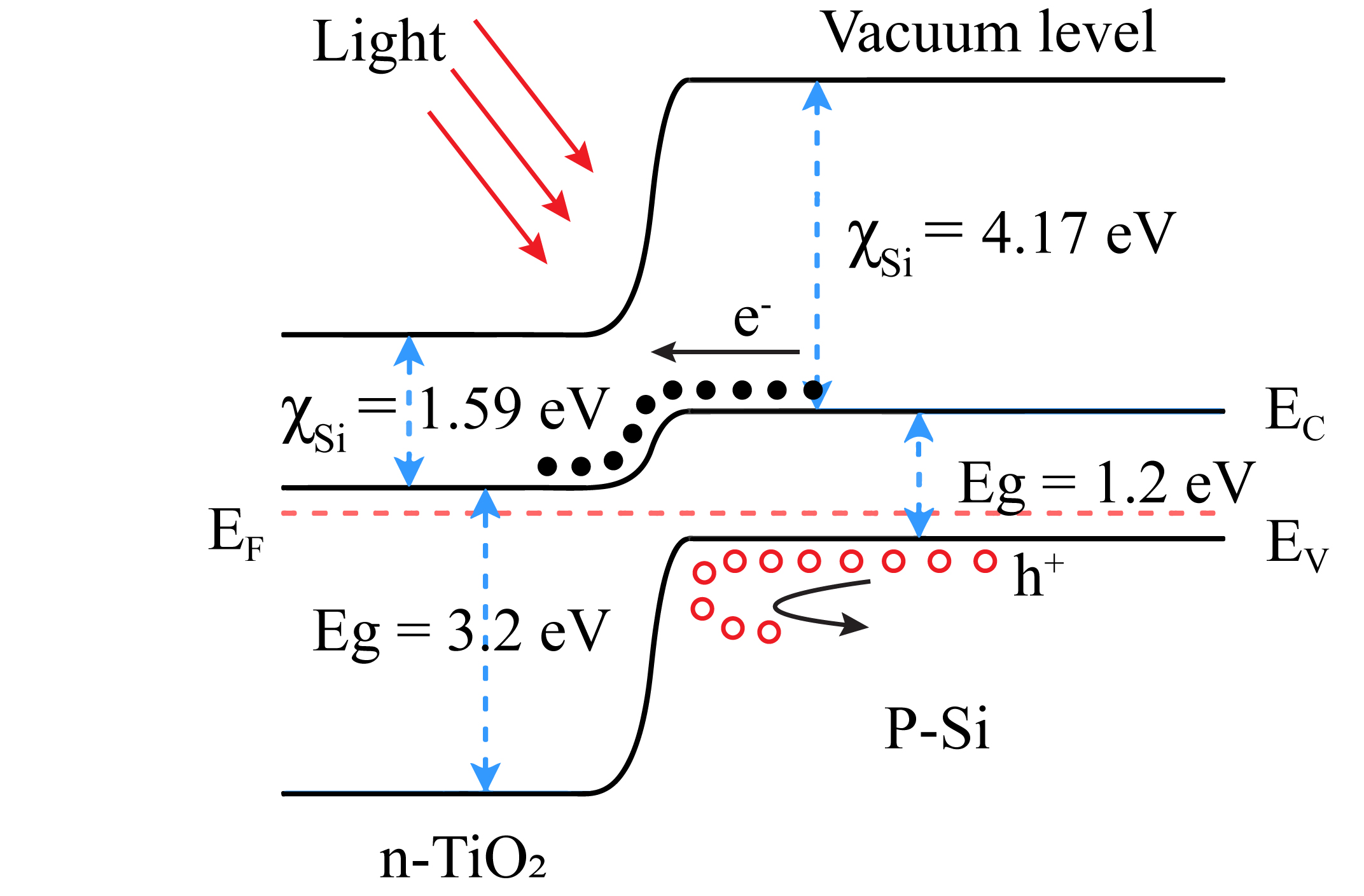}
    \caption{Energy band diagram of our proposed TFSC under illumination.}
    \label{band diagram}
\end{figure}

\begin{figure}
    \centering
    \includegraphics[width = 1\linewidth]{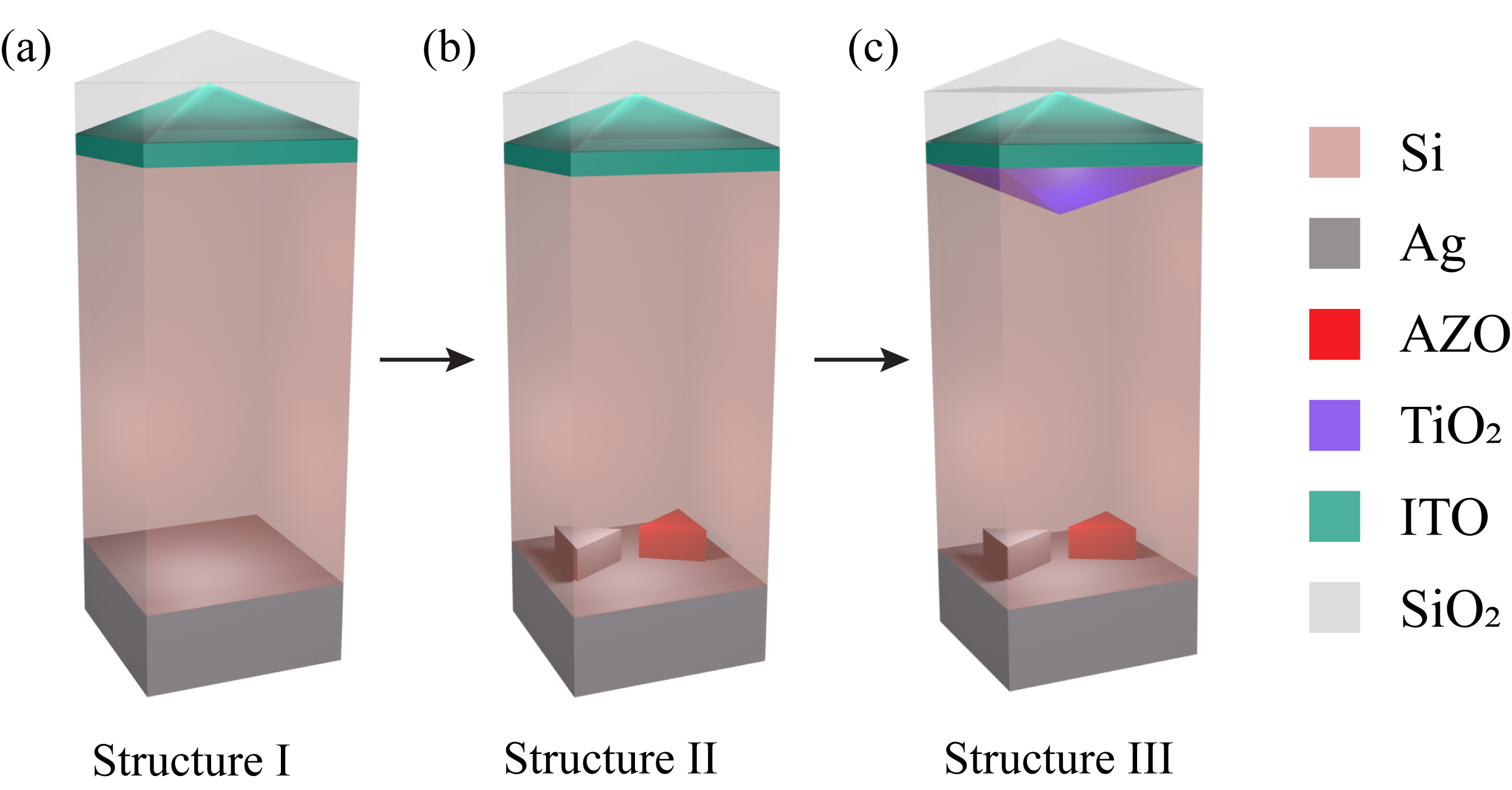}
    \caption{Schematic illustration of the (a) Structure I, (b) Structure II, (c) Structure III, respectively}
    \label{diff_structure}
\end{figure}

To explore the electrical properties of our proposed structure, we performed the device simulation in the CHARGE module of Ansys Lumerical. We doped the photoactive layer, \ce{Si} as p-type while the electron transport layer, \ce{TiO2} was doped as n-type. For both cases, the doping concentration was $1e13$. To minimize the recombination and contact resistance, a large amount of doping was applied to the region adjacent to the photoactive layer and electrode compared to the rest of the photoactive region. By considering the conductivity, we characterized the TCO materials (AZO, ITO) as metallic in our CHARGE simulation configuration. The details of the simulation configuration and material's properties have been addressed in Section S3 of the ESI. Figure~\ref{band diagram} depicts the energy band diagram of our proposed TFSC under illumination.
The transportation of the photo-generated electrons from  p-\ce{Si} to n-\ce{TiO2} was facilitated by the built-in electric field. Afterward, those electrons were collected by the electrode \ce{ITO}. Meanwhile, the valence band at the n-\ce{TiO2} exhibited a significant offset, impeding the transportation of photo-generated holes to n-\ce{TiO2}. These holes were collected by the electrode $Ag$. We considered the trap-assisted, radiative, and Auger recombination when we characterized the material properties of \ce{Si} (see \textcolor{blue}{Section 3} of \textcolor{blue}{ESI} for details). Some mathematical approaches were also utilized in charge simulation to determine current density, voltage, and other electrical performance parameters such as $J_{sc}$, $V_{oc}$, $PCE$, and $FF$ (see \textcolor{blue}{Section S3} of \textcolor{blue}{ESI} for details).\\
We divided our proposed c-\ce{Si}/\ce{TiO2} TFSC into three distinct structures to analyze the impact of various layers. Structure I was characterized by a \ce{Si} layer with BR at the bottom and a front layer comprised of \ce{ITO} and \ce{SiO2} pyramid layers while Structure II was distinguished by the inclusion of hybrid metal-dielectric NS on the BR layer in Structure I. Structure III was characterized by incorporating an electron transport layer based on \ce{TiO2} in Structure II. Figure~\ref{diff_structure} illustrates the schematic diagram of these three distinct structures.

\section{Result Analysis}\label{sec3}
\begin{figure*}
    \centering
    \includegraphics[width = 1\linewidth]{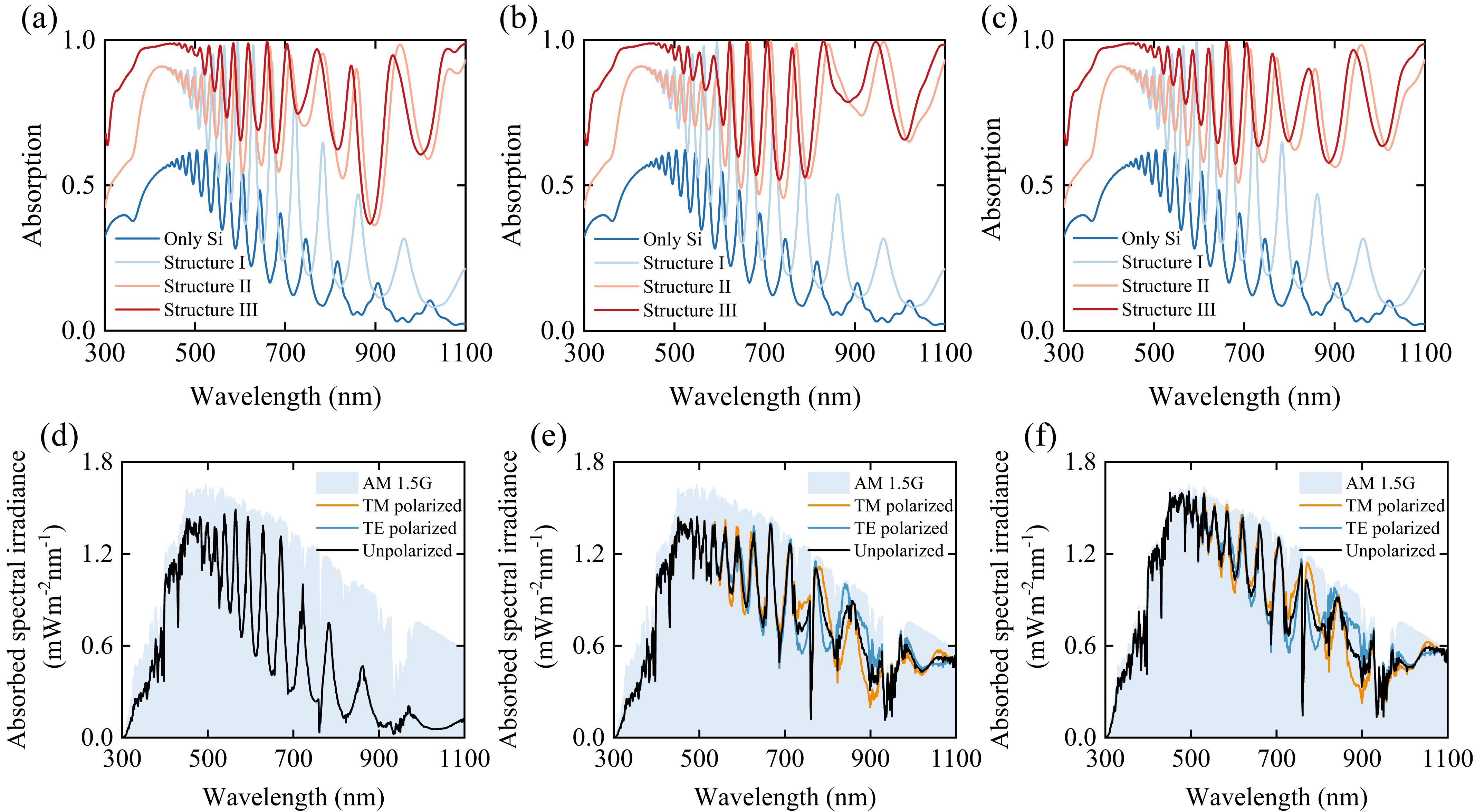}
    \caption{Absorption spectra of only \ce{Si} layer (1000 nm), Structure I, Structure II, and Structure III for (a) TM Polarized (b) TE polarized (c) Unpolarized incident light. Absorbed spectral irradiance under AM 1.5G for TM polarized, TE polarized, and unpolarized incident light of (d) Structure I (e) Structure II (f) Structure III} 
    \label{Abs_diff_struc}
\end{figure*}

\begin{figure*}
    \centering
    \includegraphics[width = 1\linewidth]{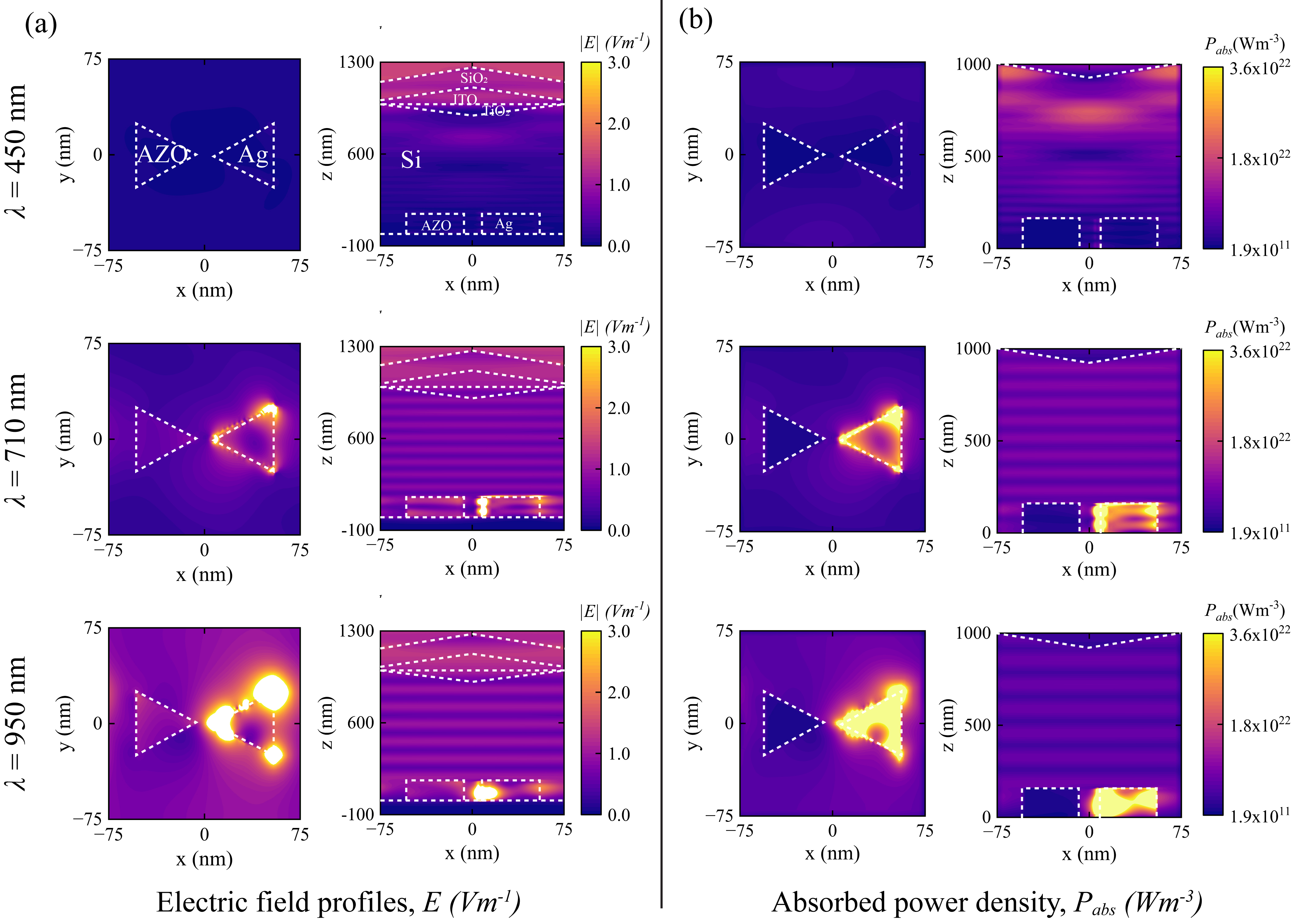}
    \caption{(a) Electric field profile (b) Absorbed power density for both xy and xz plane at $\lambda$ = 450 nm, $\lambda$ = 710 nm, $\lambda$ = 950 nm under unpolarized incident light. The wavelengths are considered from the resonance wavelengths of absorption spectra for unpolarized incident light.}
    \label{E-field}
\end{figure*}

Figures~\ref{Abs_diff_struc} (a) - (c) illustrate the absorption spectra of bulk \ce{Si} layer with thickness of 1000 nm, and the photoactive layer of Structure I, II, and III for TM polarized, TE polarized, and unpolarized incident light. The photoactive layer of Structures I and II was comprised of \ce{Si} layer, whereas the photoactive layer for Structure III included both \ce{Si} and \ce{TiO2} layers.
The bulk \ce{Si} layer exhibited an average absorption ($A_{avg}$) of 26.81\% for the incident light with a wavelength range of 300 - 1100 nm. Here, the average absorption was defined by~\cite{Nakti2023},
\begin{equation}
    A_{avg} = \frac{1}{\lambda_{max} - \lambda_{min}}\int^{\lambda_{max}}_{\lambda_{min}} A(\lambda)d\lambda
\end{equation}
Due to the symmetrical nature of Structure I, it demonstrated $A_{avg}$ of 46.37\% at the photoactive layer for both polarized and unpolarized incident light. Nevertheless, Structure I did not exhibit considerable absorption of light beyond a wavelength of 600 nm. The presence of BR at the bottom, \ce{ITO}, and \ce{SiO2} layers on top of the \ce{Si} layer in Structure I resulted in notable absorption of shorter wavelength light. Structure II improved the absorption of longer wavelength light by scattering photons that penetrated through the photoactive layer, accomplished by incorporating the HMDN on the BR. As a consequence of asymmetricity in Structure II, we observed that the $A_{avg}$ at the photoactive layer varied for TM, TE, and unpolarized incident light, with respective values of 74.37\%, 76.04\%, and 75.12\%. The absorption of shorter wavelength light was improved in Structure III with the incorporation of a \ce{TiO2}-based electron transport layer on the top surface of the photoactive layer.  Considering that the spectral irradiance intensity of the AM 1.5G spectrum is greater at shorter wavelengths compared to longer wavelengths, we decided Structure III as the final proposed structure. The $A_{avg}$ of TM, TE, and unpolarized incident light at the photoactive layer for Structure III was 82.38\%, 84.26\%, and 83.32\%, respectively. Figures~\ref{Abs_diff_struc} (d) - (f) depict the absorbed spectral irradiance under AM 1.5G for TM, TE, and unpolarized incident light of Structure I, II, and III, respectively. As mentioned earlier, Structure III revealed the highest level of absorption for any polarized light, aligning with the spectral irradiance curve of AM 1.5G with a maximum coverage of 83.32\%. Due to the unpolarized nature of sunlight, we have considered the absorbed unpolarized spectrum irradiance to obtain the generation data and associated electrical performance parameters. We incorporated the generation rate profile in Section 3 of ESI.\\
Figure~\ref{E-field} (a) and (b) illustrate the electric field profile and absorbed power density profile at the xy and xz planes for unpolarized incident light. Here, we considered three resonance wavelengths of absorption spectra, which were $\lambda$ = 450 nm, 710 nm, and 950 nm. The observed electric field and absorbed power density profiles provided insights into the impact of different layers on the absorption of incident light. 
At $\lambda$ = 450 nm, the absence of interaction between incoming photons and the NS on BR was observed. Though the Si photoactive layer possesses the capacity to absorb photons with shorter wavelengths, it is noteworthy that the top layers of our proposed structure have a substantial impact on the capture of photons with shorter wavelengths. As a result, at $\lambda$ = 450 nm, we observed a higher power density beneath the top layer, but there was a reduced power density near the NS at the bottom of the structure. Since the photon with a higher wavelength has the ability to penetrate through the photoactive layer before being absorbed, the interaction between the NS and incoming photon was observed at $\lambda$ = 710 nm and 950 nm. This scattering effect enhanced the absorption of incidence light by increasing the optical path length of incoming photons.  The observed electric field intensity close to NS  at $\lambda$ = 950 was higher than $\lambda$ = 710 nm. As a result, the power density profile followed the same path as the electric field profile. Therefore, it was obvious that the HMDN on BR had a role in absorbing light with longer wavelengths, whereas the top layer played a role in absorbing light with shorter wavelengths.

\begin{figure*}
    \centering
    \includegraphics[width = 1\linewidth]{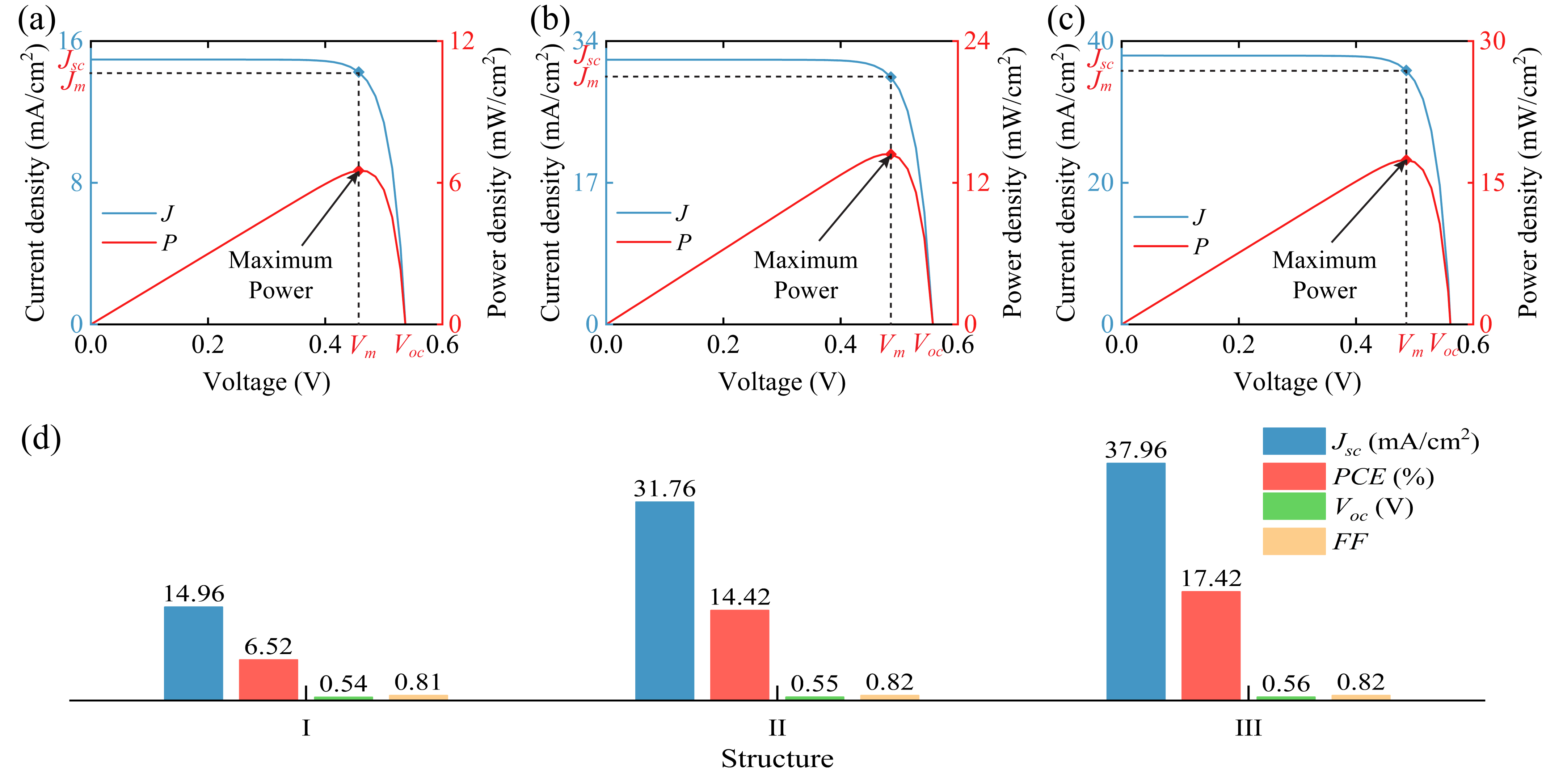}
    \caption{$J-V$ and $P-V$ characteristics of (a) structure I, (b) structure II, and (c) structure III for unpolarized incident light. The arrow sign is used to represent the maximum power point or operating point for each structure. The voltage at the highest power point is denoted as $V_m$, while the current density is denoted as $J_m$. (d) Bar chart of electrical performance parameters of structures I, II, and III for unpolarized incident light.} 
    \label{three structure}
\end{figure*}

Connecting a load between the emitter and the base of the cell creates a potential difference, resulting in the generation of a current. As previously stated, the current density, voltage, and power density were determined using several mathematical approaches. Figures~\ref{three structure} (a) - (c) shows the $J -V$ and $P - V$ characteristics of Structure I, II, and III under the illumination of an unpolarized light source.
 In the $P -V$ characteristics curve for each structure, the arrow sign represents the maximum power point, $P_{max}$ or operating point of the cell. The corresponding voltage and the current density are denoted by $V_m$ and $J_m$ respectively. The optimum load can be determined by taking the ratio of $V_m$ to $J_m$, resulting in sheet resistance~\cite{Nelson2003},
 \begin{equation}
     Optimum ~load = \frac{V_m}{J_m}
 \end{equation}
 $J_{sc}$ and $V_{oc}$ are commonly used to represent the short circuit current density and open circuit voltage, respectively. All electrical performance parameters were gradually increased from Structure I to III. Structure III yielded a maximum $J_{sc}$ of 37.96 mA/cm$^2$, $V_{oc}$ of 0.56 V, and $PCE$ of 17.42 \% among those three structures. In this particular instance, the $P - V$ characteristics curve revealed a maximum power point, $P_{max}$ of 17.42 mW/cm$^2$, corresponding to $V_m$ of 0.49 V and $J_m$ of 35.86 mA/cm$^2$. The optimal load for our proposed TFSC (Structure III) was determined to be 13.66 $\Omega$/cm$^2$. The calculated fill factor was 0.82. The $J_{sc}$ for Structure I and Structure II were found to be 14.96 mA/cm$^2$ and 31.76 mA/cm$^2$, respectively. Additionally, the $PCE$ values were determined to be 6.52 \% and 14.42 \%. The bar chart depicted in Fig~\ref{three structure} (d) provides a visual representation of the variations in electrical performance metrics between Structure I and III.

\subsection{Effect of Incident Light Polarization}
\begin{figure*}
    \centering
    \includegraphics[width = 1\linewidth]{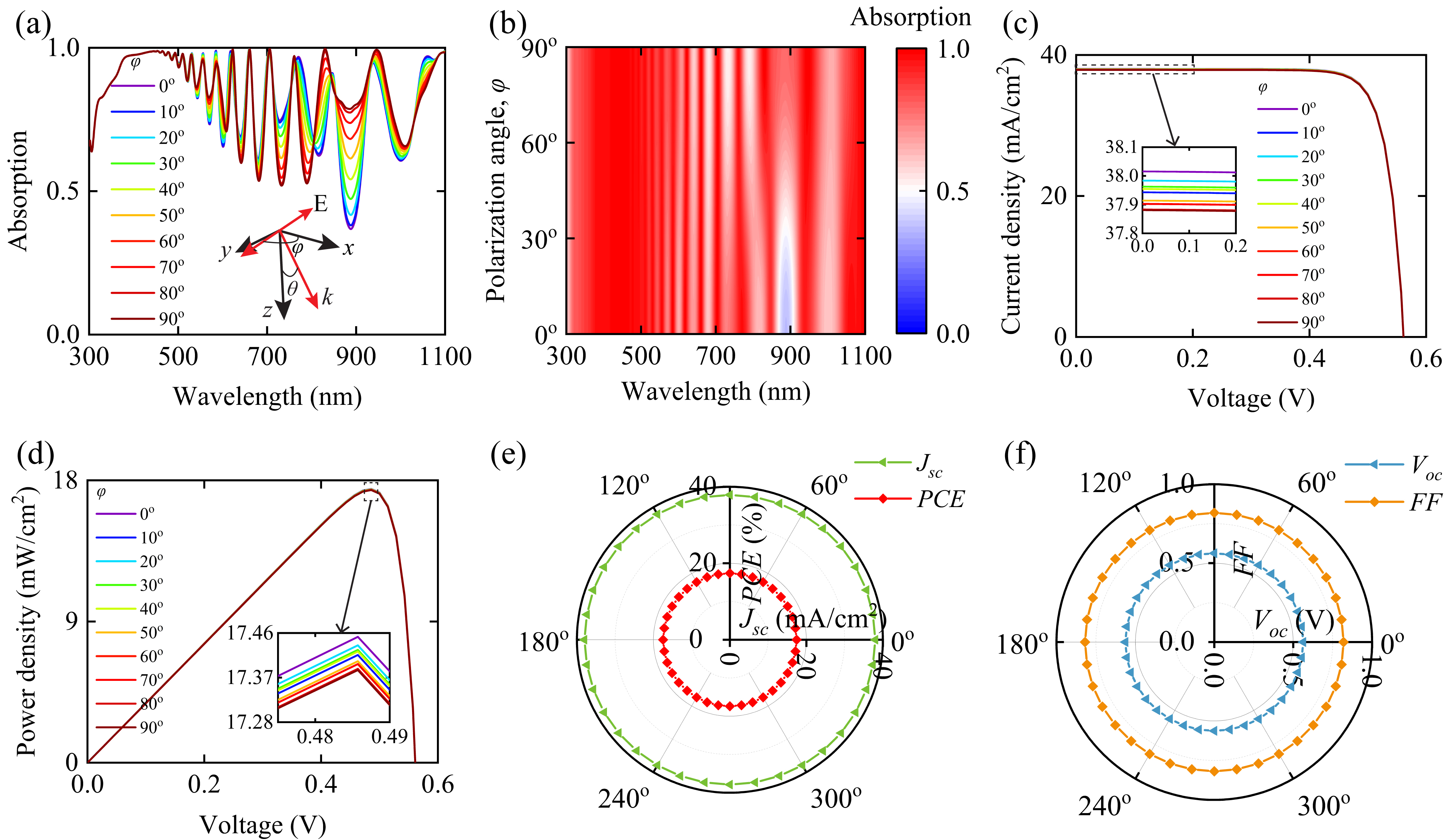}
    \caption{(a) and (b) absorption spectra of our proposed TFSC for $\varphi = 0^\circ$ to  $\varphi = 90^\circ$. The inset shows the polarization angle, $\varphi$, and the incidence angle, $\theta$ of the plane wave. (c) and (d) are the $J - V$ and $P - V$ characteristics of proposed TFSC for $\varphi = 0^\circ$ to  $\varphi = 90^\circ$. The inset shows the zoomed view of the $J - V$ and $P - V$ characteristics from where we can evaluate the $J_{sc}$ and $P_{max}$. (e) and (f) Polar plot of the electrical performance of proposed TFSC for $\varphi = 0^\circ$ to  $\varphi = 360^\circ$}
    \label{Pol angle}
\end{figure*}

In order to achieve optimal solar cell performance, it is important to maintain polarization-independent absorption capabilities. In this study, we examined both the optical and electrical performance through the utilization of a polarized light source. Figures~\ref{Pol angle} (a) and (b) depict the line and contour plots of absorption spectra in the presence of incident light with polarization angle, $\varphi$ = $0^\circ$ to $90^\circ$.
 The observed absorption spectra exhibited a variation within the wavelength range of 700 nm to 900 nm with the rotation of $\varphi$. However, this variation did not significantly impact on the $A_{avg}$. The $A_{avg}$ increased from 82.28 \% to 84.26 \% when the $\varphi$ was rotated from $0^\circ$ to $90^\circ$, indicating a relative change of 2.28 \% in $A_{avg}$. Consequently, the proposed structure demonstrated a high level of polarization angle insensitivity on the electrical performance parameters. Figures~\ref{Pol angle} (c) and (d) depict the change of $J - V$ and $P - V$ characteristics curve while varying $\varphi$ from $0^\circ$ to $90^\circ$ where the insets provide the magnified view of these characteristics curves. Based on the inset data presented in these two figures, it is obvious that the $J_{sc}$ and $P_{max}$ demonstrated a noteworthy level of tolerability over the polarization angle. As a result, the other electrical performance parameters followed the same path. The $J_{sc}$ decreased from 38.02 mA/cm$^2$ to 37.88 mA/cm$^2$, while the $P_{max}$ and $PCE$ decreased from 17.45 mW/cm$^2$ to 17.39 mW/cm$^2$ when the $\varphi$ was rotated ranging from $0^\circ$ to $90^\circ$. However, $V_{oc}$ and $FF$ maintained the constant values of 0.56 V and 0.82. Here, the $J_{sc}$ and $P_{max}$ experienced a relative change of 0.36 \%, and 0.34 \% respectively. Although the $A_{avg}$ exhibited an increment with respect to $\varphi$, the structure demonstrated a higher absorption of shorter wavelength photons at $\varphi$ = $0^\circ$ compared to $\varphi$ = $90^\circ$. As a result, both $J_{sc}$ and $PCE$ decreased when $\varphi$ rotated from $0^\circ$ to $90^\circ$. Table~\ref{tab:pol angle} enlists the electrical performance parameters of our proposed TFSC as a function of $\varphi$ of the incident light source. The polar plot of the electrical performance parameters with a variation of $\varphi$ range from $0^\circ$ to $360^\circ$ is illustrated in Figs.~\ref{Pol angle} (e) and (f). Both figures exhibited a flawless circular shape, suggesting a high level of tolerability towards incident light polarization.
\begin{table*}[]
\caption{Electrical performance parameters of our proposed TFSC for different $\varphi$ ranging from $0^\circ$ to $90^\circ$.}
\label{tab:pol angle}
\scriptsize
\begin{tabular*}{\textwidth}{@{\extracolsep\fill}ccccccccccc@{}}
\toprule
\textbf{$\varphi$} & \textbf{$0^\circ$} & \textbf{$10^\circ$} & \textbf{$20^\circ$} & \textbf{$30^\circ$} & \textbf{$40^\circ$} & \textbf{$50^\circ$} & \textbf{$60^\circ$} & \textbf{$70^\circ$} & \textbf{$80^\circ$} & \textbf{$90^\circ$} \\ \midrule
$J_{sc}$ (mA/cm$^2$) & 38.02      & 37.94       & 37.98       & 37.96       & 37.95       & 37.91       & 37.9        & 37.9        & 37.88       & 37.88       \\
$PCE$ (\%) & 17.45      & 17.42       & 17.44       & 17.43       & 17.42       & 17.4        & 17.4        & 17.4        & 17.39       & 17.39       \\
$V_{oc}$ (V)  & 0.56       & 0.56        & 0.56        & 0.56        & 0.56        & 0.56        & 0.56        & 0.56        & 0.56        & 0.56        \\
$FF$  & 0.82       & 0.82        & 0.82        & 0.82        & 0.82        & 0.82        & 0.82        & 0.82        & 0.82        & 0.82        \\ \bottomrule
\end{tabular*}
\end{table*}

\subsection{Effect of Incidence Angle}
\begin{figure*}
    \centering
    \includegraphics[width = 1\linewidth]{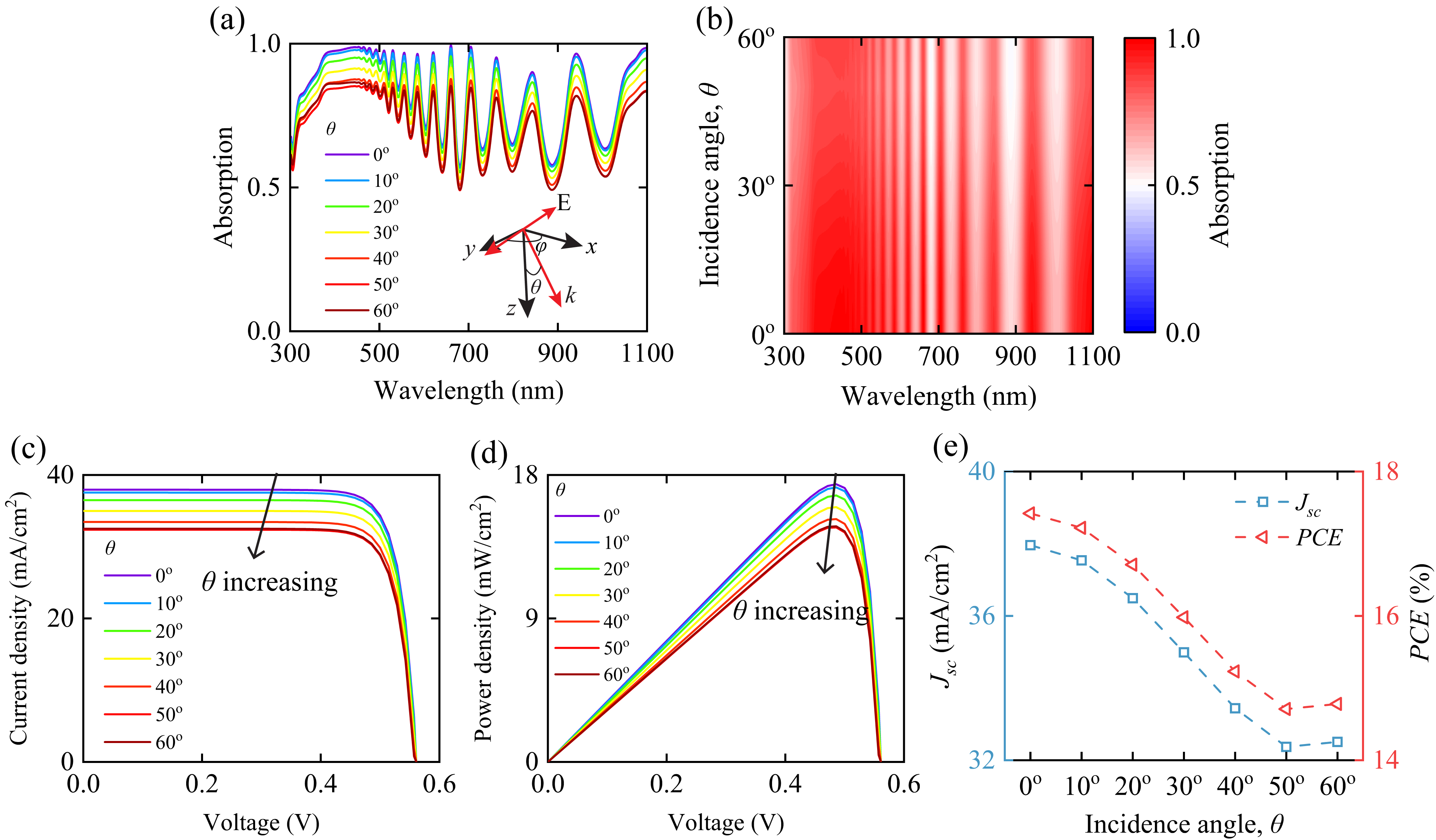}
    \caption{(a) and (b) Absorption spectra at $\theta = 0^\circ$ to $\theta = 60^\circ$ for unpolarized incident light. The inset shows the polarization angle, $\varphi$, and the incidence angle, $\theta$ of the plane wave. (c) and (d) $J - V$ and $P - V$ characteristics of our proposed structure under unpolarized incident light at $\theta = 0^\circ$ to $\theta = 60^\circ$. The arrow sign denotes the change of $J_{sc}$ and $P_{max}$ with $\theta$. (e) Comparison of Electrical parameters ($J_{sc}$ and PCE) under unpolarized incident light at $\theta = 0^\circ$ to $\theta = 60^\circ$.}
    \label{inc angle}
\end{figure*}

Throughout the day, the solar irradiance does not consistently maintain a typical incidence angle. Considering this issue, incidence angle tolerability is one of the major challenges to the performance of solar cells. In this study, we investigated the optical and electrical characteristics of our proposed TFSC by utilizing an incident light source positioned at angles ranging from $\theta$ = $0^\circ$ to $60^\circ$ along the z-axis. The line and contour plots in Fig.~\ref{inc angle} (a) and (b) depict the absorption spectra of our structure, with $\theta$ ranging from $0^\circ$ to $60^\circ$.
It is obvious that, although the absorption spectra exhibited consistent resonances, the strength of the spectra decreased. Consequently, the $A_{avg}$ decreased as $\theta$ increased. The $A_{avg}$ of 83.32\%, and 71.87 \% were observed for $\theta$ = $0^\circ$, and $60^\circ$ respectively, where a 13.74 \% of relative change was found. As a result, the electrical performance parameters also followed the same trend with $A_{avg}$ when $\theta$ was rotated. Figures~\ref{inc angle} (c) and (d) illustrate the $J - V$ and $P - V$ characteristics of our proposed structure with the variation of $\theta$. The arrow signs indicate the shifting of both characteristics curve when $\theta$ was increased. It is obvious that the $J_{sc}$ and the $P_{max}$ were decreased with the increment of $\theta$. At the normal incidence of unpolarized light, $J_{sc}$ and $P_{max}$ were determined to be 37.96 mA/cm$^2$ and 17.42 mW/cm$^2$. Until the $\theta$ = $20^\circ$, the $A_{avg}$ maintained the value above 80 \%, and $J_{sc}$ and $P_{max}$ were determined to be 36.47 mA/cm$^2$ and 16.71 mW/cm$^2$, respectively. This indicates a relative change of 3.92 \% for $J_{sc}$ and 4.07 \% for $P_{max}$.  However, when $\theta$ was rotated from $0^\circ$ to $60^\circ$, the structure exhibited a $J_{sc}$ of 32.51 mA/cm$^2$ and a $P_{max}$ of 14.78 mW/cm$^2$, although it is worth noting that the $V_{oc}$ and $FF$ values remained constant at 0.56 V and 0.82, respectively. Figure~\ref{inc angle} (e) demonstrates a comparative analysis of $J_{sc}$ and $PCE$ with the change of $\theta$ from $0^\circ$ to $60^\circ$. Moreover, the electrical performance parameters are enlisted in Table~\ref{inc angle table} with the change of $\theta$. 
\begin{table*}[]
\caption{Electrical performance parameters of our proposed TFSC for different $\theta$ ranging from $0^\circ$ to $60^\circ$}.
\label{inc angle table}
\begin{tabular*}{\textwidth}{@{\extracolsep\fill}cccccccc@{}}
\toprule
$\theta$ & $0^\circ$ & $10^\circ$ & $20^\circ$ & $30^\circ$ & $40^\circ$ & $50^\circ$ & $60^\circ$ \\ \midrule
$J_{sc}$ (mA/cm$^2$)       & 37.96      & 37.54       & 36.49       & 34.99       & 33.44       & 32.37       & 32.51       \\
$PCE$ (\%)       & 17.42      & 17.22       & 16.71       & 15.98       & 15.23       & 14.71       & 14.78       \\
$V_{oc}$ (V)      & 0.56       & 0.56        & 0.56        & 0.56        & 0.56        & 0.56        & 0.56        \\
$FF$        & 0.82       & 0.82        & 0.82        & 0.82        & 0.82        & 0.82        & 0.82        \\ \bottomrule
\end{tabular*}
\end{table*}

\subsection{Effect of Structural Parameters}
We explored the optical and electrical performance by varying some structural parameters while maintaining the other parameters at constant values. The investigations were conducted using an unpolarized incident light source. The results of these changes in structural characteristics will be addressed in the following sections.

\subsubsection{Photoactive layer thickness and Period length}
\begin{figure*}
    \centering
    \includegraphics[width = 1\linewidth]{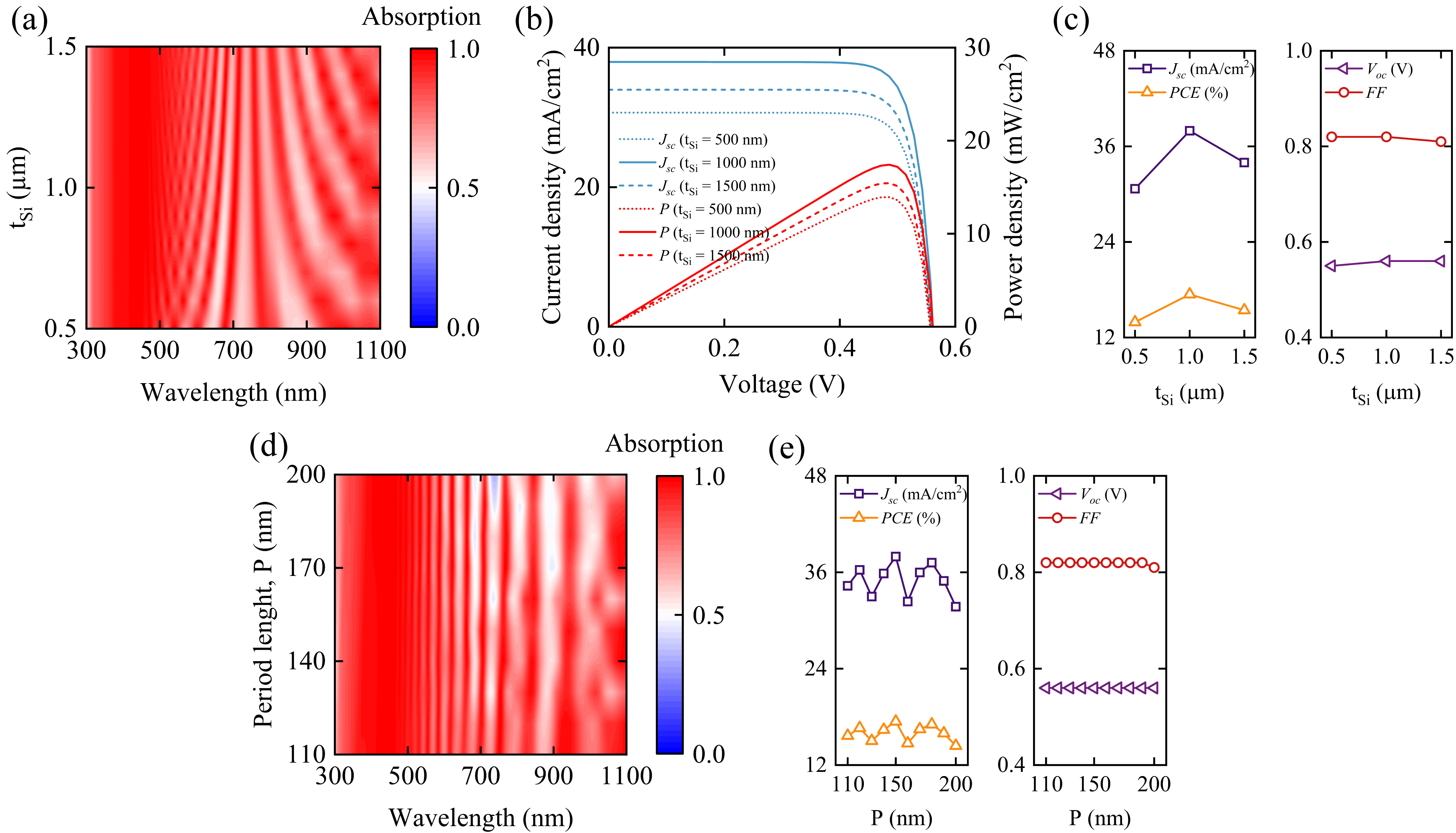}
    \caption{(a) Absorption spectra for t$_{\ce{Si}}$ = 500 nm to 1500 nm, while the other structural parameters were kept constant. (b) $J - V$ and $P - V$ characteristics for t$_{\ce{Si}}$ = 500 nm, 1000 nm, and 1500 nm (c) Comparison of electrical performance parameters for t$_{\ce{Si}}$ = 500 nm, 1000 nm, and 1500 nm (d) Absorption spectra for P = 110 nm to 200 nm while the other structural parameters were kept constant. (e) Comparison of electrical performance parameters for  P = 110 nm to 200 nm. Here, we considered unpolarized incidence light for both cases.} 
    \label{Si thick and period}
\end{figure*}

Since we considered c-\ce{Si} as our photoactive layer, the thickness of \ce{Si} layer is one of the major concerns because \ce{Si} is an indirect bandgap material. In our proposed structure, we utilized a 1000 nm of c-\ce{Si} layer to generate photocarriers while the other structural parameters were optimized to absorb the highest amount of unpolarized incident light. Here, we investigated the optical performance for t$_{\ce{Si}}$ = 500 nm to 1500 nm while the other structural parameters were kept constant. Figure~\ref{Si thick and period} (a) presents the absorption spectra for t$_{\ce{Si}}$ = 500 nm to 1500 nm under unpolarized incident light.
 With the increment of the t$_{\ce{Si}}$, the number of resonances as well as the quality factor were seen to be increased. Moreover, we found that $A_{avg}$ was increased with the increment of t$_{\ce{Si}}$. The $A_{avg}$ increased from 82.36 \% to 83.90 \% when the t$_{\ce{Si}}$ was increased from 500 nm to 1500 nm. However, since we considered t$_{\ce{Si}}$ of 1000 nm in our structure and optimized doping concentration, and doping area, we found better electrical performance for t$_{\ce{Si}}$ =1000 nm than t$_{\ce{Si}}$ = 1500 nm. Figure~\ref{Si thick and period} (b) depicts the $J - V$ and $P - V$ characteristics for t$_{\ce{Si}}$ = 500 nm, 1000 nm, and 1500 nm. We have found the $J_{sc}$ of 30.68 mA/cm$^2$, 37.96 mA/cm$^2$, and 33.97 mA/cm$^2$ and $P_{max}$ of 13.93 mW/cm$^2$, 17.42 mW/cm$^2$, and 15.43 mW/cm$^2$ for t$_{\ce{Si}}$ of 500 nm, 1000 nm, and 1500 nm, respectively. Figure~\ref{Si thick and period} (c) illustrates a comparative study of electrical performance parameters for different values of t$_{\ce{Si}}$. Our proposed structure, with a 1000 nm \ce{Si} photoactive layer, clearly demonstrated superior performance compared to alternative structures.

 Furthermore, a study was conducted to determine the impact of the period length, P of our structure on both optical and electrical performance. We varied the P from 110 nm to 200 nm while the other structural parameters were kept constant. The variation of P was initiated from 110 nm to contain the NS within this period length. Figure~\ref{Si thick and period} (d) illustrates the absorption spectra with varying P where an irregularity in resonance as well as $A_{avg}$ was observed. The electrical performance parameters also displayed the same trend, like $A_{avg}$. Figure~\ref{Si thick and period} (e) depicts the comparative analysis of the electrical performance parameters for P = 110 nm to 200 nm. Three peak values were observed for $J_{sc}$ and $PCE$ at P = 120 nm, 150 nm, and 180 nm which included our proposed period length. At P = 120 nm and 180 nm, the values of $J_{sc}$ were determined to be 36.3 mA/cm$^2$ and 37.2 mA/cm$^2$, while the values of $PCE$ were found to be 16.61 \% and 17.06 \%. However, for all instances, the $V_{oc}$ and $FF$ remained constant which were 0.56 V and 0.82. 
 
\subsubsection{Height and Distance of Nanostructure}
To scatter the longer wavelength photons that penetrate through the photoactive layer, we employed a pair of hybrid metal-dielectric isosceles triangle nanostructures that were organized in a periodic pattern on the back reflector. The dimensions and alignment of NS dictate the interaction between light and matter, as well as the absorption of light. In this study, we investigated the impact of the thickness, t$_{\ce{ns}}$ and the distance between a pair of NS, d$_{\ce{ns}}$ on both optical and electrical performance.

\begin{figure*}
    \centering
    \includegraphics[width = 0.65\linewidth]{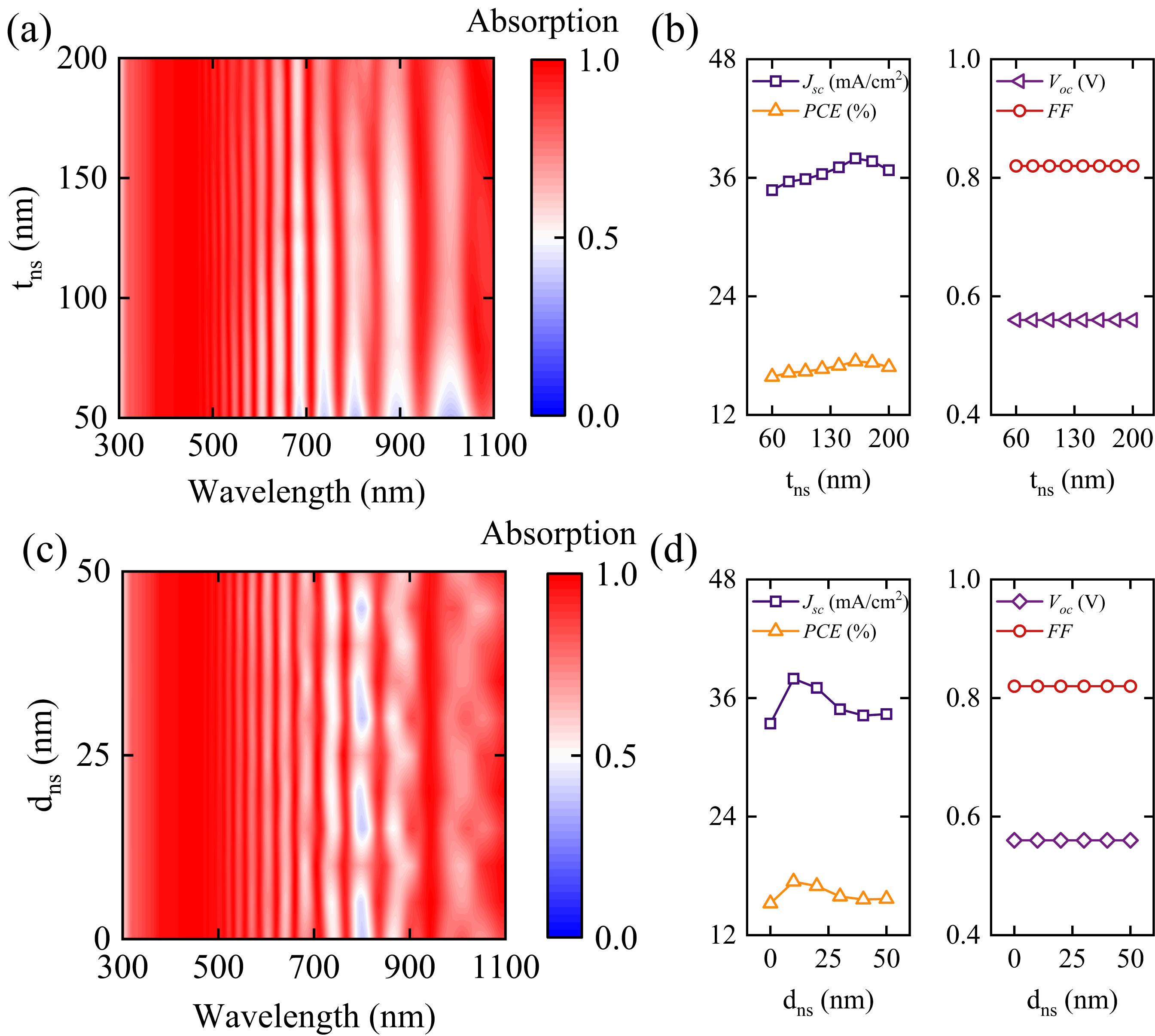}
    \caption{(a) Absorption spectra for t$_{\ce{ns}}$ = 50 to 200 nm for unpolarized incidence light (b) Comparison graph of electrical performance parameters for t$_{\ce{ns}}$ = 60 nm to 200 nm with an increment of 20 nm (c) Absorption spectra for d$_{\ce{ns}}$ = 0 to 50 nm for unpolarized incidence light (d)  Comparison graph of electrical performance parameters for d$_{\ce{ns}}$ = 0 nm to 50 nm with an increment of 10 nm.}
    \label{thic dist of NS}
\end{figure*}

To investigate the optical performance, the thickness of both NS varied from 50 nm to 200 nm, however, their side lengths and the distance between the pair of NS were maintained at a constant value. Moreover, we maintained the other structural parameters constant as before. Figure~\ref{thic dist of NS} (a) depicts the absorption spectra varying the t$_{\ce{ns}}$ where it is obvious that the resonance wavelengths remained the same with the change of t$_{\ce{ns}}$. Nevertheless, both the intensity of the resonance and the $A_{avg}$ exhibited a consistent increase until t$_{\ce{ns}}$ reached 160 nm. Following the t$_{\ce{ns}}$ of 160 nm, the $A_{avg}$ exhibited a decreasing trend as the t$_{\ce{ns}}$ increased. Due to the direct proportionality between the generated photocurrent and the absorption of photons, we have proposed the structure with t$_{\ce{ns}}$ of 160 nm. Figure~\ref{thic dist of NS} (b) depicts a comparative analysis of the electrical performance parameters of our proposed structure, with the t$_{\ce{ns}}$ ranging from 60 nm to 200 nm where the $J_{sc}$ and $PCE$ exhibited a similar trend as $A_{avg}$. At t$_{\ce{ns}}$ = 60 nm, $J_{sc}$ and $PCE$ were determined to be 34.75 mA/cm$^2$ and 15.86 \% while at t$_{\ce{ns}}$ = 200 nm, $J_{sc}$, and $PCE$ were determined to be 36.77 mA/cm$^2$ and 16.85 \%. Both the $J_{sc}$ and $PCE$ demonstrated an upward trend as the t$_{\ce{ns}}$ increased from 60 nm to 160 nm. At t$_{\ce{ns}}$ = 160 nm, the $J_{sc}$, and $PCE$ were determined to be 37.96 mA/cm$^2$ and 17.42 \%, respectively. However, for further increase of t$_{\ce{ns}}$, both values were reduced. The values of $V_{oc}$ and $FF$ remained consistent while the t$_{\ce{ns}}$ varied, with values of 0.56 V and 0.82, respectively. The structure we proposed, with t$_{\ce{ns}}$ of 160 nm, demonstrated superior performance compared to the other structures.

The distance between the HMDN pair was one of the major parameters to enhance the absorption. We investigated the impact of the distance between the NS, d$_{\ce{ns}}$ ranging from 0 nm to 50 nm on optical and electrical performance. Figure~\ref{thic dist of NS} (c) depicts the absorption spectra varying the d$_{\ce{ns}}$. Among all the distances, we found that d$_{\ce{ns}}$ with 10 nm exhibited the highest $A_{avg}$. As a result, at d$_{\ce{ns}}$ = 10nm, we found optimal electrical performance, $J_{sc}$ of 37.96 mA/cm$^2$ and $PCE$ of 17.42 \%. A comparative analysis of electrical performance parameters is illustrated in fig.~\ref{thic dist of NS} (d). With increasing the d$_{\ce{ns}}$ from 10 nm, both the $A_{avg}$ and electrical performances demonstrated a downward trend. However, the $V_{oc}$ and $FF$ remained constant with the change of d$_{\ce{ns}}$. The NS with d$_{\ce{ns}}$ = 0 exhibited the lowest electrical performance where the $J_{sc}$ and $PCE$ were determined to be 33.42 mA/cm$^2$ and 15.22 \% respectively. 

\subsection{Photovoltaic performance under non-isothermal conditions}
In order to incorporate the impact of self-heating effects, the charge transport solver was expanded to consider the influence of a temperature gradient by modifying the drift-diffusion equations (see \textcolor{blue}{Section S3} of \textcolor{blue}{ESI} for details).

\begin{figure*}
    \centering
    \includegraphics[width = 1\linewidth]{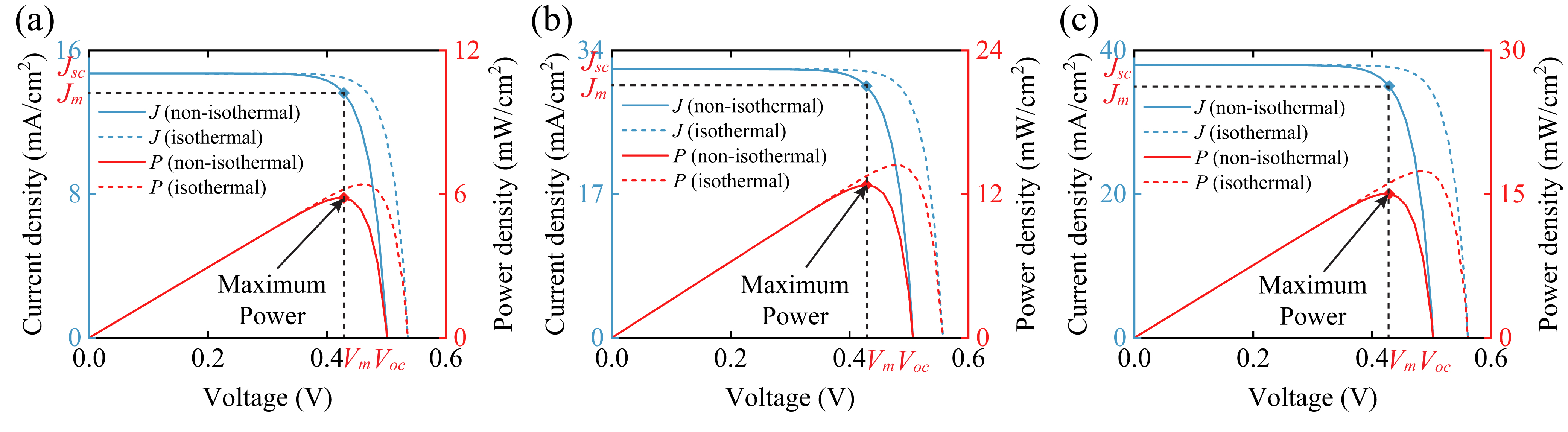}
    \caption{$J -V$ and $P - V$ characteristics under isothermal and non-isothermal conditions for (a) structure I (b) structure II and (c) structure III. The arrow sign is used to represent the maximum power point or operating point for each structure. The voltage at the highest power point is denoted as $V_m$, while the current density is denoted as $J_m$.}
    \label{nonisothermal}
\end{figure*}

We conducted the investigation of self-heating on Structure I, II, and III at non-isothermal conditions. The temperature profiles of the photoactive region of these structures are illustrated in ESI (see \textcolor{blue}{Section S3} of \textcolor{blue}{ESI} for details). Figure~\ref{nonisothermal} demonstrates the $J - V$ and $P - V$ characteristics of Structure I, II, and III under isothermal and non-isothermal conditions. The arrow sign is used to represent the maximum power point or operating point for each structure at non-isothermal conditions. The voltage at the highest power point is denoted as $V_m$, while the current density is denoted as $J_m$.
In non-isothermal conditions, the $J_{sc}$ experienced a negligible change while a notable change of $V_{oc}$ was observed. All three structures exhibited a reduced $V_{oc}$ of 0.5 V since the $V_{oc}$ is directly proportional to the bandgap. As a result, the maximum power point or operating point was changed for all three structures. We measured the $P_{max}$ of 5.86 mW/cm$^2$, 12.76 mW/cm$^2$, and 15.02 mW/cm$^2$ for Structure I, II, and III respectively. For our proposed structure, the $J_m$ and $V_m$ were determined to be 35.05 mA/cm$^2$ and 0.43 V. In this case, the optimum load was 12.27 $\Omega$/cm$^2$. A 13.77 \% relative change of $PCE$ was observed at non-isothermal conditions. We determined the $FF$ of 0.79 for all three structures under non-isothermal conditions. Table~\ref{noniso table} enlists the electrical performance parameters for Structure I, II, and III under isothermal and non-isothermal conditions. 
\begin{table*}[]
\caption{Electrical performance parameters of our proposed TFSC under isothermal and non-isothermal conditions.}
\label{noniso table}
\begin{tabular*}{\textwidth}{@{\extracolsep\fill}cccccc@{}}
\toprule
                                & Structure & $J_{sc}$ (mA/cm$^2$)   & $PCE$ (\%)   & $V_{oc}$ (V)  & $FF$   \\ \midrule
\multirow{3}{*}{Isothermal}     & I         & 14.96 & 6.52  & 0.54 & 0.81 \\
                                & II        & 31.76 & 14.42 & 0.55 & 0.82 \\
                                & III       & 37.96 & 17.42 & 0.56 & 0.82 \\ \midrule
\multirow{3}{*}{Non-isothermal} & I         & 14.72 & 5.84  & 0.5  & 0.79 \\
                                & II        & 31.77 & 12.76 & 0.5  & 0.79 \\
                                & III       & 37.97 & 15.02 & 0.5  & 0.79 \\ \midrule
\end{tabular*}
\end{table*}

\subsection{Effect of different nanostructure's material}
\begin{figure*}
    \centering
    \includegraphics[width = 0.8\linewidth]{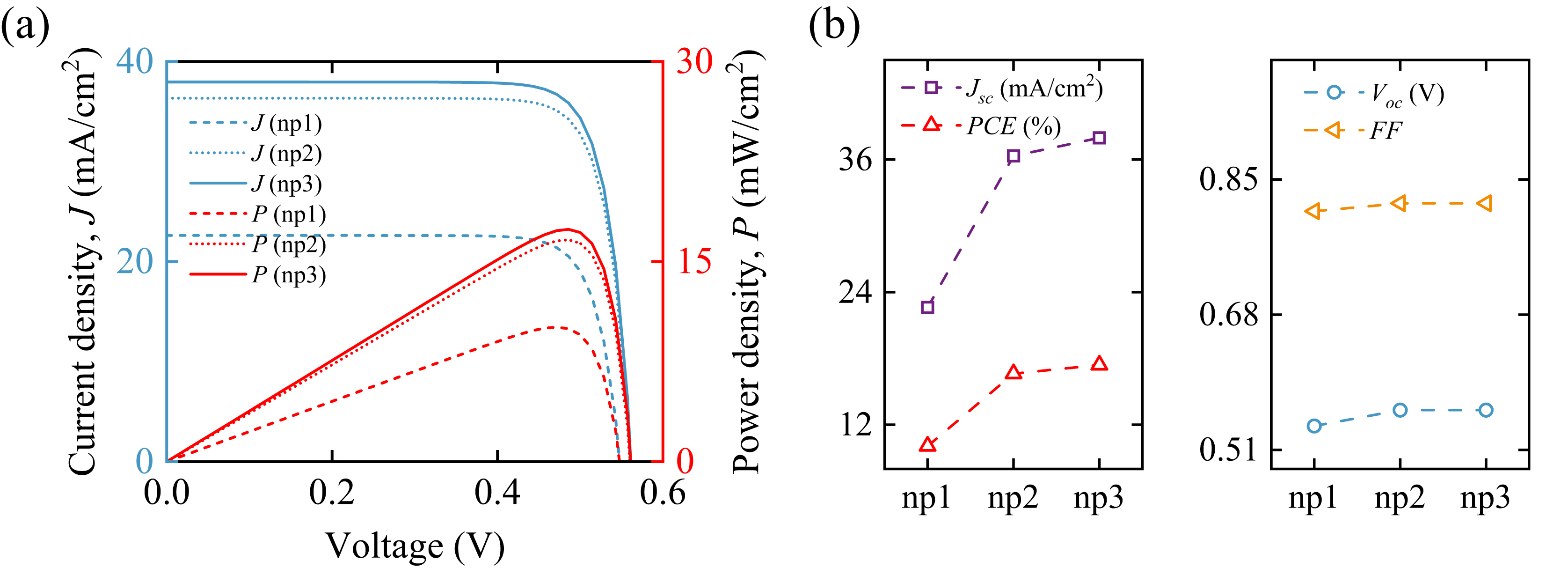}
    \caption{$J - V$ and $P -V$ characteristics for different NS pair np1, np2, and np3 comprising with \ce{AZO}-\ce{AZO}, \ce{Ag}-\ce{Ag}, and \ce{Ag}-\ce{AZO} respectively (b) Comparison of electrical performance parameters of our proposed structure with np1, np2, and np3.}
    \label{ns material}
\end{figure*}

We conducted a study to investigate the impact of the material of NS while maintaining all other structural parameters constant. We have individually analyzed the electrical performance of the structure with three pairs of NS. The first NS pair (np1) consisted of the dielectric material \ce{AZO}, while the second NS pair (np2) consisted of the metal \ce{Ag}. The third pair of NS (np3) was our proposed HMDN pair comprised of \ce{Ag} and \ce{AZO}. Figure~\ref{ns material} (a) illustrates the $J - V$ and $P - V$ characteristics for these three pairs of NS. The measured $J_{sc}$ and $V_{oc}$ for np1 were 22.61 mA/cm$^2$ and 0.54 V, respectively. Both the np2 and np3 demonstrated $V_{oc}$ of 0.56 V while $J_{sc}$ of 36.33 mA/cm$^2$ and 37.96 mA/cm$^2$ were found for np2 and np3. Figure~\ref{ns material}  (b) depicts the comparative analysis of electrical performance among these three NS pairs. Among the three NS pairs, np3 demonstrated the highest $PCE$ of 17.42 \% while for np1 and np2, the $PCE$ were determined to be 10.08 \% and 16.63 \% respectively. The HMDN enhanced the $PCE$ of metallic NS by 4.54 \%. The $FF$ of 0.82 was observed for both np2 and np3, whereas np1 had an $FF$ of 0.81. It is obvious that our proposed HMDN-based TFSC demonstrates the best electrical performance.
 \section{Comparative Analysis of Electrical Performances}
Table~\ref{comp} demonstrates a comparative study of electrical performance among our proposed structures and previously reported structures. Both Tabrizi \textit{et al.} and Mohsin \textit{et al.} proposed two solar structures with 3 $\mu$m of \ce{Si} photoactive layer where they utilized the operating wavelength of 300 – 1100 nm~\cite{Tabrizi2020, Mohsin2023}. A periodic pyramid structure comprised of \ce{TiO2} was utilized as an anti-reflection layer in both structures while Tabrizi \textit{et al.} and Mohsin \textit{et al.} utilized spherical NS and pyramid-shaped NS respectively. The $J_{sc}$ and $V_{oc}$ were determined to be 31.57 mA/cm$^2$ and 0.626 V for the structure by Tabrizi \textit{et al.}, on the other hand, 24.8723 mA/cm$^2$ and 0.923 V were found for the structure by Mohsin \textit{et al.}. A dual metallic plasmonic back reflector-based thin film solar cell was proposed by Shahabi \textit{et al.} where $J_{sc}$ and $PCE$ were determined to be 19.86 mA/cm$^2$ and 15.65 \% respectively~\cite{Shahabi2023}. Saravanan \textit{et al.} reported an amorphous Si-based solar cell with an operating wavelength of 300-1200 nm where they observed a $J_{sc}$ of 16.25 mA/cm$^2$~\cite{Saravanan2022}. Moreover, Abdi \textit{et al.} reported a spherical nanoparticle-based solar structure with a 300 nm photoactive layer and 400-1100 nm operating wavelength~\cite{Abdi2022}. They observed a $J_{sc}$ of 18.057 mA/cm$^2$ and $PCE$ of 13.03 \% with $V_{oc}$ of 0.8352 V. Meanwhile, Subhan \textit{et al.} reported a Si solar cell with a photoactive layer thickness of less than 300 nm~\cite{Subhan2020}. They incorporated bi-metallic nanograting to enhance the absorption within the wavelength range of 700- 1100 nm. They reported the $J_{sc}$ of 22.3 mA/cm$^2$ for their proposed structure. Heidarzadeh \textit{et al.} proposed an Al-Ag cylindrical NS-based structure with a 1 $\mu$m Si photoactive layer~\cite{Heidarzadeh2022}. The $J_{sc}$ value of 16.57 mA/cm$^2$ was reported for their proposed structure, which operated through the wavelength range of 300-1100 nm. Zhao \textit{et al.} also proposed a solar structure with the same photoactive layer thickness and operating wavelength range~\cite{Zhao2021}. They introduced \ce{TiO2} as an electron transport layer and Ag-based spherical NS to enhance absorption. Their finding indicated a $J_{sc}$ of 32.81 mA/cm$^2$, $PCE$ of 14.16 \%, and $V_{oc}$ of 0.53 V. On the other hand, we proposed a solar structure with improved electrical performance by incorporating the HMDN at the back reflector. We observed a $J_{sc}$ of 37.96 mA/cm$^2$ and, a $V_{oc}$ of 0.56 V. The structure exhibited a $PCE$ of 17.42 \% within the wavelength range of 300-1100 nm. Consequently, it is obvious that our proposed structure demonstrates remarkably improved electrical performance in comparison to other structures.
\begin{table*}[]
\caption{Comparative analysis of electrical performance parameter between our structure and previously reported structures}
\label{comp}
\scriptsize
\begin{tabular*}{\textwidth}{@{\extracolsep\fill}cccccccc@{}}
\toprule
\begin{tabular}[c]{@{}c@{}}Photoactive\\ Layer\end{tabular} & \begin{tabular}[c]{@{}c@{}}Thickness\\ ($\mu$m)\end{tabular}& \begin{tabular}[c]{@{}c@{}}Operating\\ Wavelength (nm)\end{tabular} & \begin{tabular}[c]{@{}c@{}}$J_{sc}$\\ (mA/cm$^2$)\end{tabular} & \begin{tabular}[c]{@{}c@{}}$V_{oc}$\\ (V)\end{tabular}   & \begin{tabular}[c]{@{}c@{}}$PCE$\\ (\%)\end{tabular}    & $FF$     & DOI         \\ \midrule
Si                  & 3         & 300 - 1100               & 31.57        & 0.62    & 16.18      & 0.82   & \cite{Tabrizi2020}           \\
Si                  & 3         & 300 - 1100               & 24.87      & 0.92     & 18.58    & 0.80 & \cite{Mohsin2023}           \\
Si                  &    --      &    --                  & 19.86        &      --     & 15.65      &   --     & \cite{Shahabi2023}           \\
Si                  &    --       & 300 - 1200           & 16.25        &     --      &     --       & --       & \cite{Saravanan2022}           \\
Si                  & 0.3  & 400 - 1100               & 18.05      & 0.83    & 13.03      & 0.73   & \cite{Abdi2022}           \\
Si                  & 0.2       & 700 - 1100               & 22.3         &     --      &     --       &    --    & \cite{Subhan2020}           \\
Si                  & 1         & 300 - 1100               & 16.57        &     --      &     --       &    --    & \cite{Heidarzadeh2022}          \\
Si                  & 1         & 300 - 1100               & 32.81        & 0.53      & 14.16      & 0.81   & \cite{Zhao2021}           \\
Si                  & 1         & 300 - 1100               & 37.96        & 0.56      & 17.42      & 0.82   & This   work \\ \bottomrule
\end{tabular*}
\end{table*}

\section{Conclusions}
In conclusion, the absorption of the incoming incident light was enhanced by both the top layers and the NS at the back reflector. The incorporation of the \ce{TiO2} inverted pyramid layer, in combination with the \ce{ITO} and \ce{SiO2}-based pyramid layers, enhanced the shorter wavelength light absorption by increasing the optical path and facilitating the coupling of incoming light in photonic mode. The triangular HMDN enhanced the longer wavelength light absorption by scattering and coupling the light in SPR mode. Consequently, the proposed structure exhibited an average absorption of 83.32 \% for the AM 1.5G solar spectrum within the wavelength range of 300 – 1100 nm. The electric field profile and absorbed power density profile effectively demonstrated the distinctive contributions of each layer in the absorption of light at shorter and longer wavelengths. The efficiency of our proposed structure exhibited a maximum relative change of 0.34 \% when subjected to a polarized light with an angle of $0^\circ$ to $90^\circ$ which ensures the insensitivity of our proposed structure against polarized light. We conducted a comprehensive analysis of both the optical and electrical performance over incident angles ranging from $0^\circ$ to $60^\circ$. Moreover, the photovoltaic performance was investigated by changing several structural parameters, where our proposed structure exhibited significantly improved electrical performance. The electrical performance was also investigated in non-isothermal conditions by including the effect of the self-heating properties, where a 13.77 \% relative change of $PCE$ was observed. In addition, the comparative analysis, which was carried out among the dielectric nanostructures, metallic nanostructures, and hybrid metal-dielectric nanostructures, revealed that the hybrid metal-dielectric nanostructures demonstrated more effective performance compared to the other configurations. Our study will pave the way to designing c-Si thin film solar cells with hybrid metal-dielectric nanostructures, which will be beneficial for the progression of photovoltaic performance.

\section*{CRediT authorship contribution statement}
\textbf{Soikot Sarkar:} Conceptualization, Investigation, Methodology, Software, Visualization, Writing – original draft. \textbf{Sajid Muhaimin Choudhury:} Funding Allocation, Supervision,  Writing – review \& editing.

\section*{Declaration of Competing Interest}
The authors declare that they have no known competing financial
interests or personal relationships that could have appeared to influence
the work reported in this paper. 

\section*{Data availability}
The data underlying the findings in this article is currently unavailable to the public. Data are obtained upon reasonable request from the authors.

\section*{Acknowledgement}
Soikot Sarkar and Sajid Muhaimin Choudhury convey their sincere appreciation to the Department of Electrical and Electronic Engineering at Bangladesh University of Engineering and Technology (BUET) for providing the Ansys Lumerical software and the required support.

\bibliographystyle{elsarticle-num} 
\bibliography{ref}


\clearpage
\vspace{10mm}
\textbf{Electronic Supplementary Information (ESI)}
\section{Structural Parameters Optimization}
\begin{figure*}[htbp]
    \centering
    \includegraphics[width = 1 \linewidth]{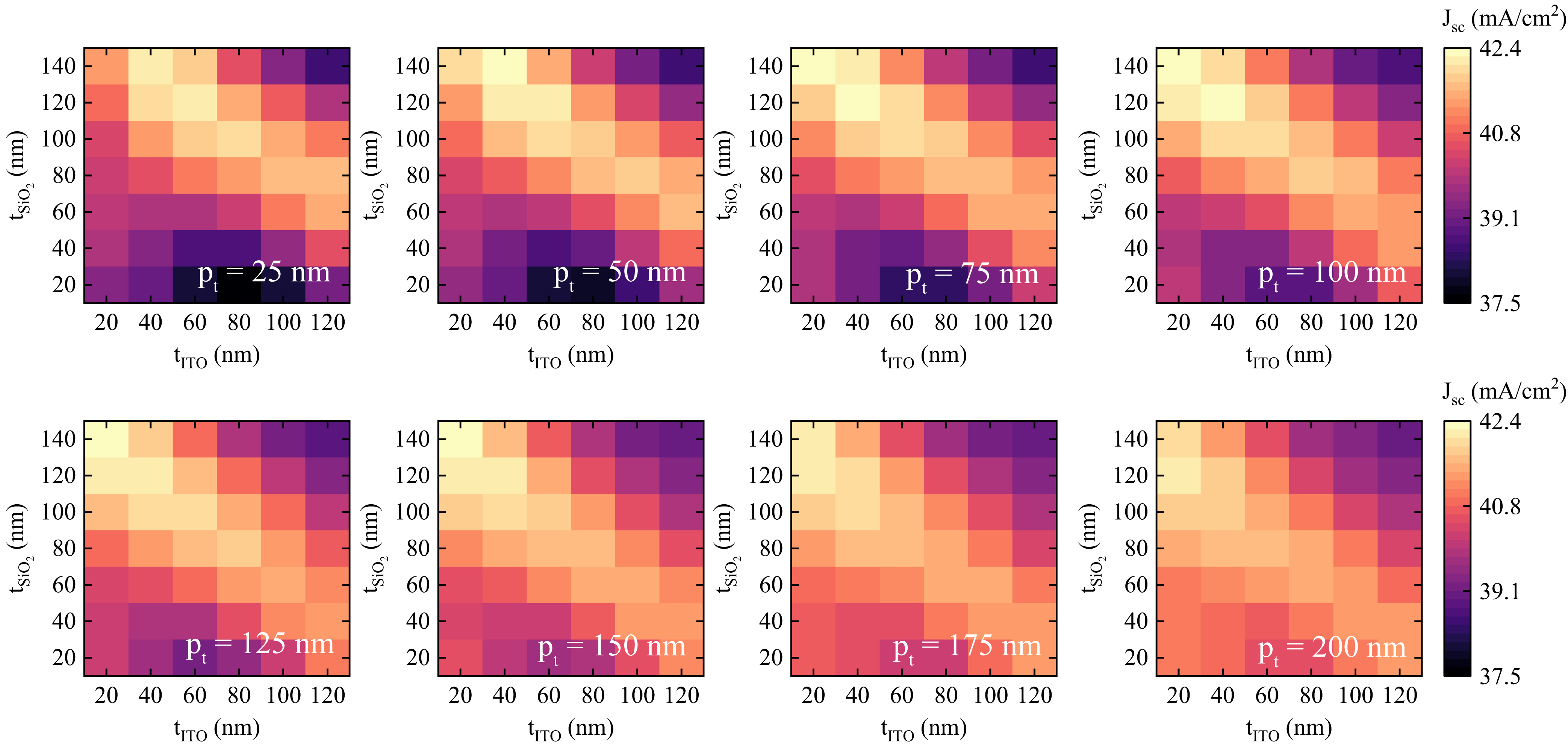}
    \caption{Short-circuit current density determined by varying various structural parameters of the front layers in optical simulation}
    \label{fig:front layer}
\end{figure*}
\begin{figure*}[htbp]
    \centering
    \includegraphics[width = 1 \linewidth]{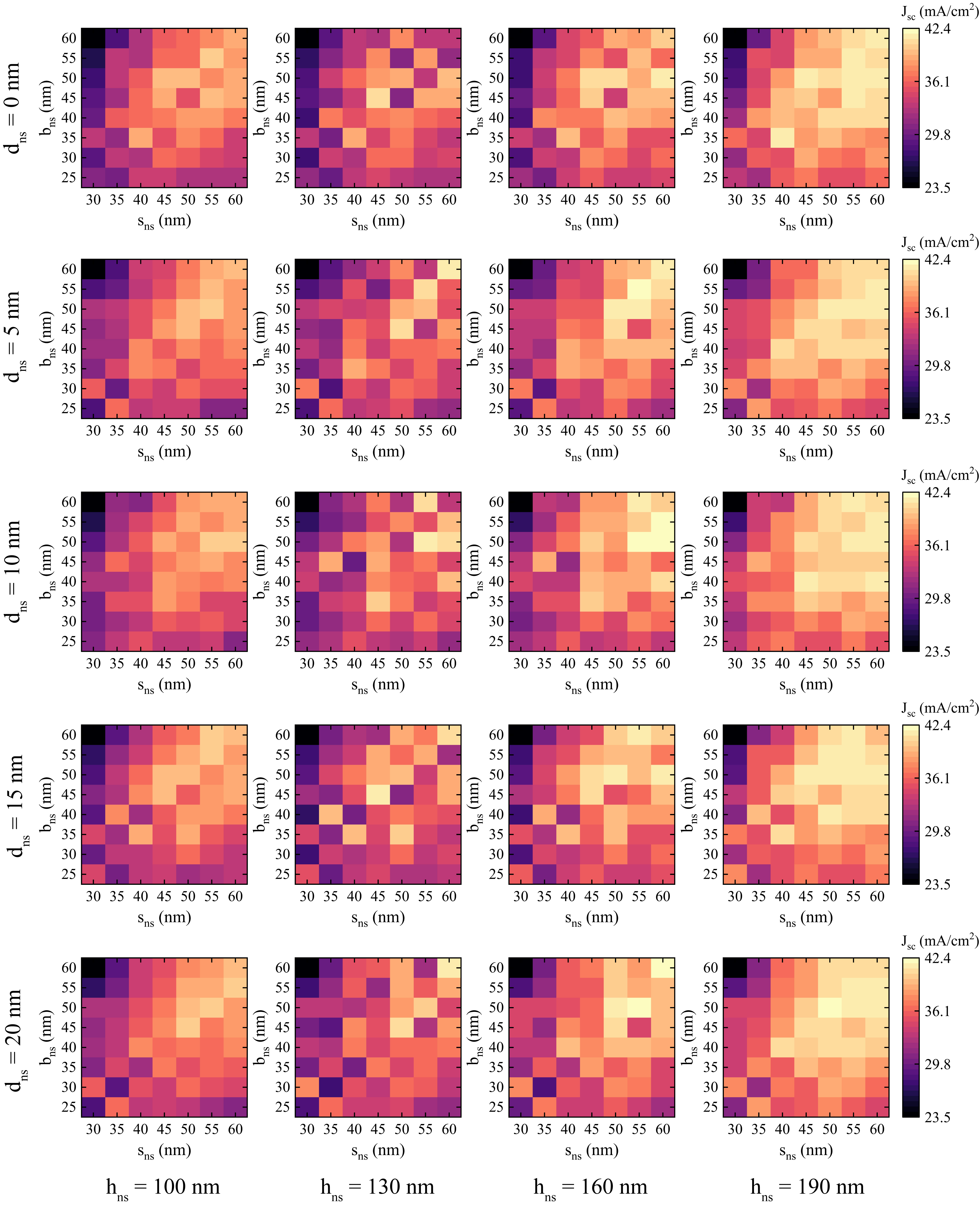}
    \caption{Short-circuit current density determined by varying various structural parameters of nanostructures in optical simulation}
    \label{fig:ns}
\end{figure*}
The structural parameters were optimized by utilizing the short circuit current density, $J_{sc}$ obtained from the optical simulation, using the following equation,

\begin{equation}
    J_{sc} = e\int\frac{\lambda}{hc}QE(\lambda)I_{AM1.5}(\lambda)d\lambda
\end{equation}
We optimized the thicknesses of the front layers and the dimensions of the nanostructures separately. We have conducted these operations several times to get an optimized performance. Figure~\ref{fig:front layer} and \ref{fig:ns} represent the short circuit current density determined by optical simulation at the final iteration. 

\section{Material properties}
\begin{figure*}[htbp]
    \centering
    \includegraphics[width = 0.85 \linewidth]{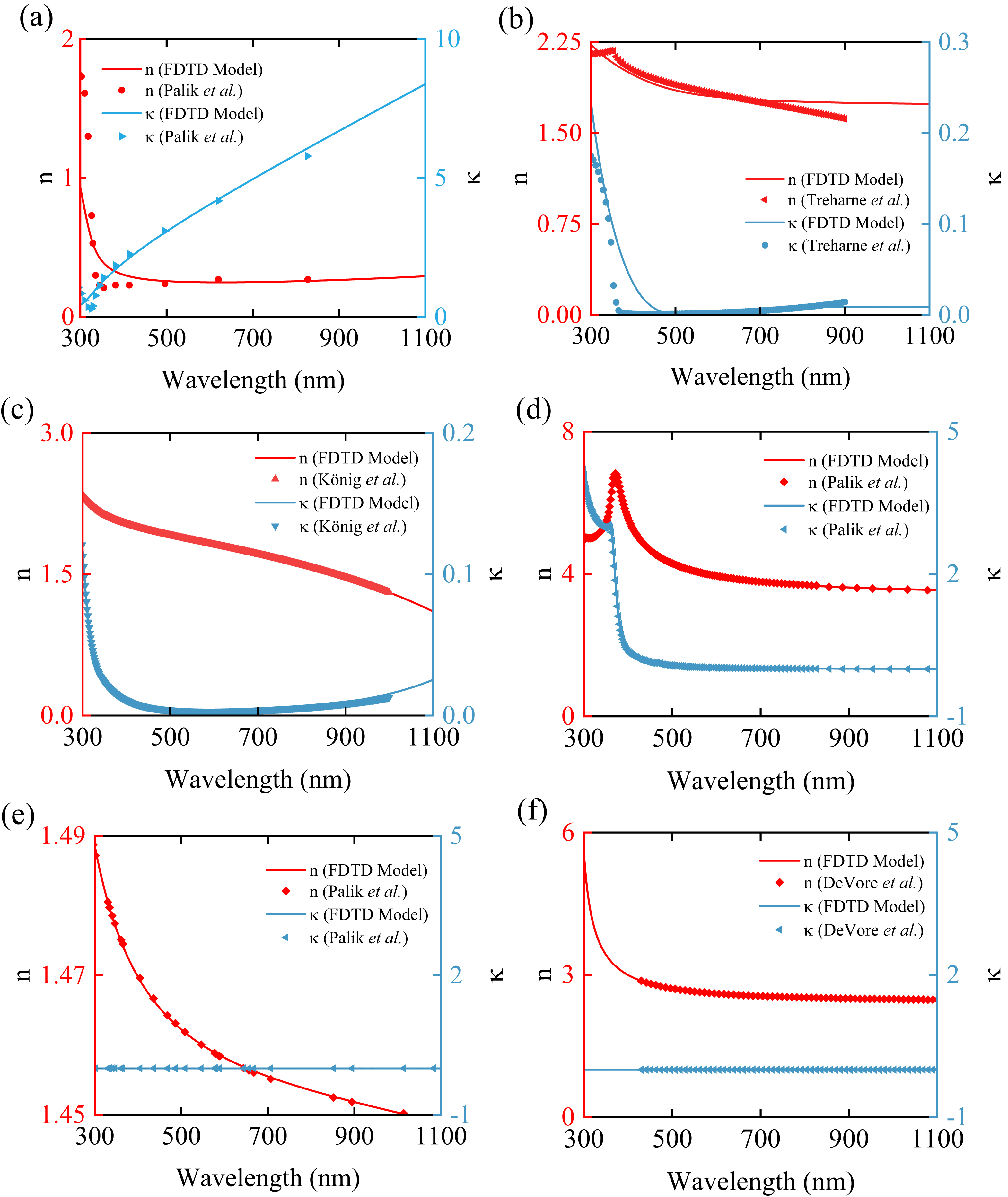}
    \caption{The refractive index (n) and extinction coefficient ($\kappa$) spectra of (a) \ce{Ag}, (b) \ce{AZO}, (c) \ce{ITO}, (d) \ce{Si}, (e) \ce{SiO2}, and (f) \ce{TiO2}.}
    \label{material data}
\end{figure*}

Figure~\ref{material data} illustrates both the adopted and FDTD fitted optical properties of \ce{Ag}, \ce{AZO}, \ce{ITO}, \ce{Si}, \ce{SiO2}, and \ce{TiO2} which include both the refractive index (n), and the extinction coefficient ($\kappa$). The speed of the light is controlled by the value of n, while $\kappa$ determines how much the light will be scattered and absorbed. 

\end{document}